\documentclass[prb,aps,twocolumn,amsmath,amssymb, superscriptaddress]{revtex4}
\usepackage[pdftex]{graphicx}
 \usepackage{amsmath}
\usepackage[T1]{fontenc}
\usepackage{amssymb}
\usepackage{amsfonts}
\usepackage{flushend}
 \usepackage{amsmath} 
\usepackage{lipsum}
\usepackage{amsfonts} 
\usepackage{amssymb, mathrsfs}
\usepackage{braket}
\usepackage{graphicx} 
\usepackage{subfigure}
\usepackage{bbm}
\def\beq{\begin{equation}}
\def\eeq{\end{equation}}
\def\bsp{\begin{split}}
\def\esp{\end{split}}
\def\bea{\begin{eqnarray}}
\def\eea{\end{eqnarray}}
\def\ba{\begin{array}}
\def\ea{\end{array}}

\def\dg{\dagger}

\def\lb{\left(}
\def\rb{\right)}

\def\l.{\left.}
\def\r.{\right.}

\def\ra{\rangle}
\def\la{\langle}

\def\bo{\bold{k}}

\begin{document}

\title{Magnon Hall effect  in AB-stacked bilayer honeycomb quantum magnets}
\author{S. A. Owerre}
\affiliation{Perimeter Institute for Theoretical Physics, 31 Caroline St. N., Waterloo, Ontario N2L 2Y5, Canada.}
\email{sowerre@perimeterinstitute.ca}
\affiliation{African Institute for Mathematical Sciences, 6 Melrose Road, Muizenberg, Cape Town 7945, South Africa.}

\begin{abstract}
Motivated by the fact that many bilayer quantum magnets occur in nature,  we generalize the study of thermal Hall transports of spin excitations to bilayer magnetic systems.   It is shown that bilayer magnetic systems can be   coupled either ferromagnetically or antiferromagnetically.  We study both scenarios on the honeycomb lattice and show that the system realizes topologically nontrivial magnon bands induced by alternating next-nearest-neighbour Dzyaloshinsky-Moriya  interaction (DMI). As a result, the bilayer system realizes both magnon Hall effect and magnon spin Nernst effect. We show that antiferromagnetically coupled layers differ from ferromagnetically coupled layers  by a sign change in the   conductivities as the magnetic field is reversed. Furthermore,   Chern number protected magnon edge states are observed and propagate in the same direction on the top and bottom layers in ferromagnetically coupled layers, whereas the magnon edge states propagate in opposite directions for antiferromagnetically coupled layers.
\end{abstract}
\maketitle

\section{Introduction}

Magnons are the collective excitations of ordered quantum magnets such as ferromagnets or antiferromagnets. In quantum magnetic systems that lack inversion symmetry, the  DMI  (spin-orbit coupling)\cite{dm} is present  and leads to  chiral magnons with nontrivial topological properties \cite{alex0,alex1,zhh,shin,shin1, alex1a} similar to electronic systems \cite{yu6,yu7,fdm}.  As magnons are uncharged (neutral) quasiparticles, they do not experience a magnetic field in the Lorentz force as in the electronic version of Hall effect. Instead,  they exhibit a thermal version of Hall effect in which a temperature gradient $\boldsymbol \nabla T$ transports a heat current $\bold J_Q$ \cite{zhh, alex1, alex0, alex7, alex4, alex44}.    The DMI generates a nonzero Berry curvature given by \cite{alex2} $\boldsymbol{\Omega}(\bold k)={\nabla}_{\bf k}\times \bold{A}(\bold k)$, where $\bold{A}(\bold k)$ is a DMI dependent vector potential.  The Berry curvature  acts as an effective magnetic field by altering the propagation of magnons in the system, thus leads to thermal Hall effect dubbed magnon Hall effect  \cite{alex1, alex0}, as well as magnon spin Nernst effect \cite{alex7}.   These two phenomena are characterized by two conductivities ---  transverse thermal conductivity $\kappa_{xy}$ and transverse spin Nernst conductivity $\alpha_{xy}^s$. They are both  directly related to the Berry curvature of the magnon bulk bands reminiscent of Hall conductivity in electronic  systems \cite{thou}. However, in contrast to  electronic  systems, there is no completely filled bands in bosonic systems, so each magnon band contributes a term to $\kappa_{xy}$ and $\alpha_{xy}^s$, and the Chern number of the system simply leads to protected chiral magnon edge states.

The first experimental  realization of magnon Hall effect has been  reported   in three-dimensional (3D)  pyrochlore ferromagnetic insulators  Lu$_2$V$_2$O$_7$, Ho$_2$V$_2$O$_7$, and In$_2$Mn$_2$O$_7$ \cite{alex1,alex1a}. Quite recently,  magnon Hall effect has been observed in 2D kagome magnet Cu(1-3, bdc) \cite{alex6, alex6a}. Both theory and experiment show that the topology of the system  leads to a sign change in   $\kappa_{xy}$  as a function of temperature or magnetic field on the kagome lattice \cite{alex4,alex6, alex6a} and a sign change in $\kappa_{xy}$   as the magnetic field is reversed on the  pyrochlore lattice \cite{alex1, alex1a}.  These experimental observations have propelled  a possibility of numerous experimentally accessible 2D ferromagnets in different lattices  that exhibit nontrivial topological spin excitations. In this regard,  magnon Hall transports have been proposed on a single-layer  Lieb ferromagnet \cite{xc} with three magnon bulk bands, in which a sign change was also observed in $\kappa_{xy}$. The honeycomb ferromagnet is special as it applies to experimentally accessible graphene sheet \cite{yu6,yu7, cas}. It also forms an  example of nontrivial topological energy bands in electronic systems \cite{fdm}. In the context of magnon Hall transports,   the author has recently shown that nontrivial magnon bands and magnon Hall effect could be accessible in a single-layer honeycomb quantum ferromagnets, in which a DMI is allowed by the alternating triangular plaquettes of the  next-nearest-neighbour  sites \cite{sol, sol1}. For the honeycomb lattice, $\kappa_{xy}$ shows no  sign change for all parameters of the system \cite{sol1}. Recent study has shown that the proposed honeycomb ferromagnetic system also exhibits spin Nernst effect \cite{kkim}.  It has been previously  shown that an interfacial contact between   two  magnon insulators can lead to nontrivial  transport properties   on the kagome ferromagnet \cite{mook}. 

Motivated  by these theoretical and experimental realizations of thermal Hall effects of spin excitations and the experimental realizations of bilayer honeycomb-lattice quantum  magnetic compounds  such as Na$_3$Cu$_2$SbO$_6$ \cite{aat1}, $\beta$-Cu$_2$V$_2$O$_7$ \cite{aat}, and  A$_2$IrO$_3$ (A= Na, Li) \cite{aat0,aat00}, we study thermal Hall transports of magnons in bilayer honeycomb magnetic systems.  Generalization to other bilayer systems  is straightforward. We show that two layers of magnon insulators with the same DMI can be coupled  either ferromagnetically or antiferromagnetically. In the former, we show that the presence of an alternating next-nearest-neighbour DMI  generates Berry curvatures  and dissipationless magnon edge states,  which propagate in the same direction on the top and the bottom layers of the bilayer system.  We compute the thermal Hall and spin Nernst conductivities. 
For the latter case, the spins on the upper\slash lower layer point in opposite direction to those on the lower\slash upper layer, hence the interlayer couplings become antiferromagnetic. In this case, we show that magnon edge states  propagate in opposite directions. As a result, thermal Hall and spin Nernst conductivities also show a change sign as the magnetic field is reversed. This result is interesting because it is what is seen experimentally in many ferromagnetic insulators \cite{alex1,alex1a}.    As  thermal Hall transports of magnons await experimental observation on the honeycomb lattice, our results  open new avenue to search for 2D and 3D bilayer systems with nonzero thermal Hall transports. 

\section{Bilayer magnon insulator}
\begin{figure}[ht]
\centering
\includegraphics[width=.75\linewidth]{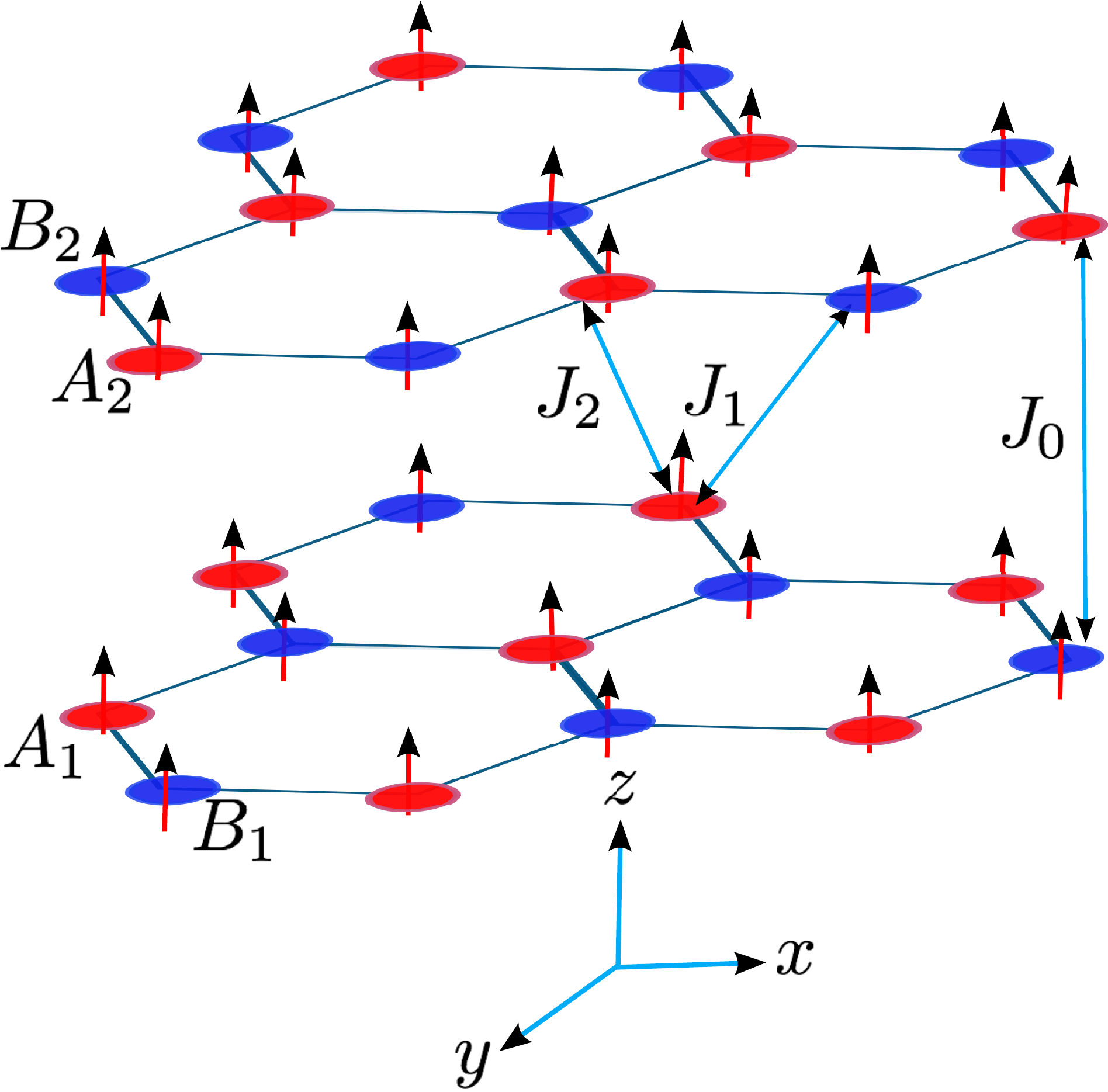}
\caption{Color online.  Lattice structure of ferromagnetically coupled bilayer honeycomb ferromagnet with various hopping parameters. For antiferromagnetically coupled layers, the spins on the upper layer point in opposite direction to those on the lower layer.}
\label{lattice1}
\end{figure}

Bilayer magnon insulators consist of two single-layer of magnon insulators coupled either ferromagnetically or antiferromagnetically as shown in Fig.~\ref{lattice1}. The Hamiltonian is governed by \begin{align}
H&= H_{FM,\tau}+ H_{DMI,\tau}+H_{ext,\tau}+H_{int.},
\label{h}
\end{align}
where, $H_{\tau}$ represents the single-layer Hamiltonians for the top $\tau=T$ and bottom $\tau=B$ layers respectively, $H_{DMI,\tau}$ represents the DMI interactions on both layers, $H_{ext,\tau}$ is the external magnetic field on each layer,  and $H_{int}$ is the interlayer couplings between them. For the honeycomb lattice, they are given by
\begin{align}
&H_{FM,\tau}=-J\sum_{\la i, j\ra}{\bf S}_{i}^\tau\cdot{\bf S}_{j}^\tau\\ &H_{DMI,\tau}=D\sum_{\la \la i,j\ra\ra} \nu_{ij}\bold{\hat z}\cdot{\bf S}_{i}^\tau\times{\bf S}_{j}^\tau,\label{model}\\ &H_{ext, \tau}=-h\sum_{i}S_{i,z}^\tau,\\& H_{int.}=-\sum_{ i\in T, j\in B; \alpha}J_{\alpha}{\bf S}_{i}\cdot{\bf S}_{j},
\label{h2}
\end{align}
where ${\bf S}_{i}$ is the spin moment at site $i$, $J>0$ is a nearest-neighbour (NN) ferromagnetic interaction on each layer, and $D$ is the magnitude of the  DMI which is allowed by the next-nearest-neighbour (NNN) triangular plaquettes on the honeycomb lattice, where  $\nu_{ij}=\pm 1$ represents hopping from left to right and vice versa on the NNN sites.  The Zeeman magnetic field is $h$ in units of $g\mu_B$. The interlayer  NN interactions  $J_{\alpha}>0$ denotes all the possible interlayer couplings depicted in Fig.~\ref{lattice1}.

\section{Magnon bands}
   In this section, we present the band structure of the bilayer honeycomb magnetic system using the linearized HP boson representation of the spin operators \cite{HP}, which is valid at low-temperature when few magnons are thermally excited. In fact, this formalism has been employed  frequently in the study of thermal Hall transports of magnons in ferromagnetic systems \cite{zhh, alex0, alex1, alex2, alex4}. It is also supported in recent experimental realizations \cite{alex6, alex6a, alex1}.   

 \subsection{Ferromagnetically coupled layers}
We first start with ferromagnetically coupled layers, $J_\alpha >0$.  Using the linearized HP transformation, the magnon tight binding hopping model  is given by
\begin{align}
H_{\tau}&=v_s^\prime\sum_{i} b_{i}^\dagger b_{i} -v_s\sum_{\la ij\ra}b_{i}^\dagger b_{j} - v_D\sum_{\la \la ij\ra\ra}i\nu_{ij}b_{i}^\dagger b_{j},\label{hp3}\\
H_{int.}&= \sum_{i\in T, j\in B; \alpha}v_{\alpha}[( b_{i}^\dagger b_{i}+ b_{j}^\dagger b_{j}) -( b_{i}^\dagger b_{j}+ b_{j}^\dagger b_{i})],
\label{hpp3}
\end{align}
where  $v_s^\prime=zv_s + h$,  $v_s(v_D)= JS(DS)$, and  $z=3$ is coordination number of the lattice.   The interlayer coupling is $v_{\alpha}=J_{\alpha}S$. 
Similar to AB-stacked bilayer graphene\cite{mcc, mcc1,mcc2,mcc3}  there are three kinds of interlayer couplings as shown in Fig.~\ref{lattice1}. We identify the bottom layer with two sublattices labeled $A_1$ and $B_1$, and the top layer with $A_2$ and $B_2$. The exchange interaction $J_0$ couples $B_1$ and $A_2$, $J_1$ couples $A_1$ and $B_2$, and $J_2$ couples $A_1$ and $A_2$, $B_1$ and $B_2$.  In Fourier space  the momentum space Hamiltonian is given by $ H=\sum_{\bold k}\psi^\dagger_{\bold k}\cdot \mathcal{H}_{FM}(\bold k)\cdot\psi_{\bold k},$  with the basis $\psi^\dagger_{\bold k}= (b_{\bold{k}A_1}^{\dagger},\thinspace b_{\bold{k} B_1}^{\dagger},b_{\bold{k} A_2}^{\dagger},\thinspace b_{\bold{k} B_2}^{\dagger})$, where the Bogoliubov Hamiltonian is given by
\begin{figure}[ht]
\centering
\includegraphics[width=4in]{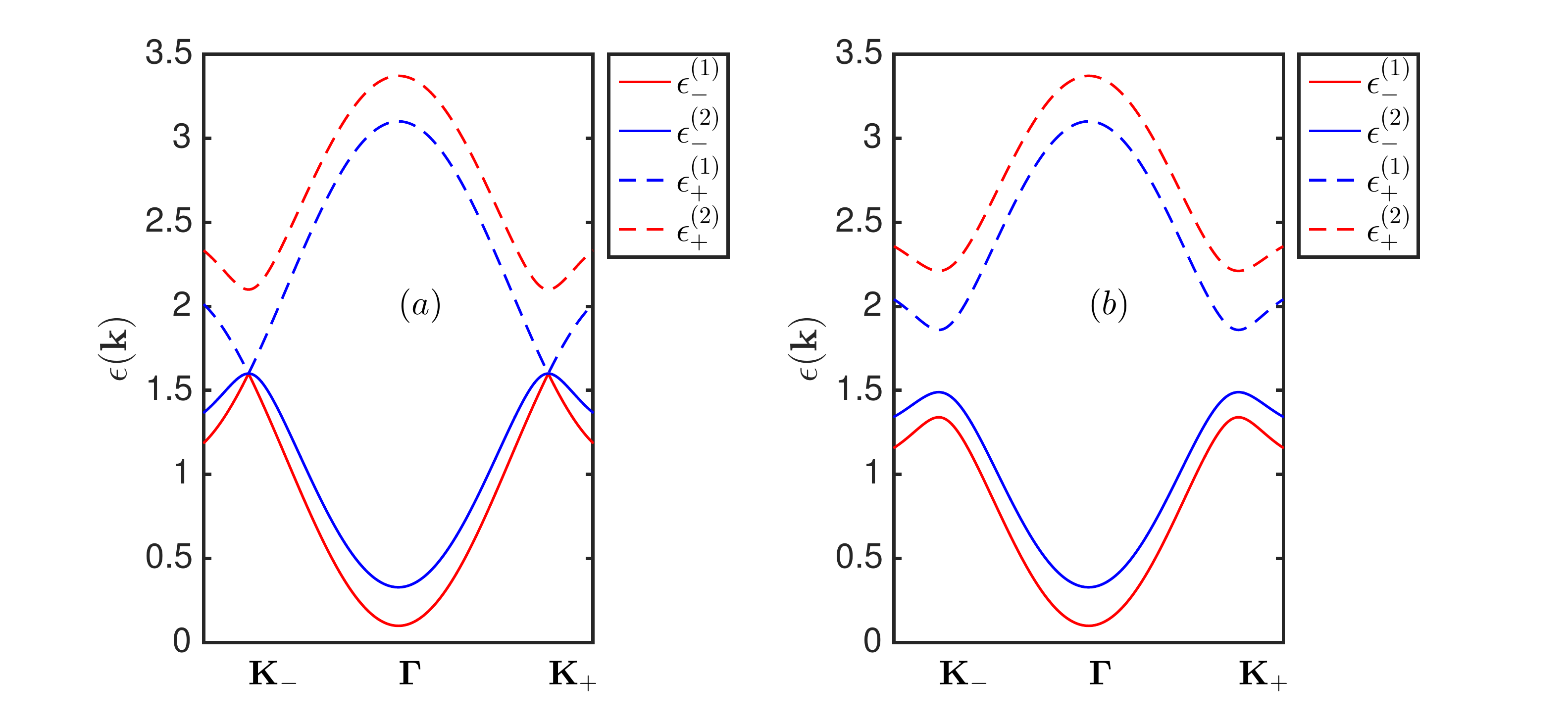}
\caption{Color online. The magnon bulk bands of the spin-$1/2$ bilayer ferromagnet along $k_y=0$ at  $v_s=0.5,~h=0.1,v_0=0.25,~v_1=v_2=0$:  $(a)$ $v_{D}=0.0$ $(b)$  $v_{D}=0.05$.}
\label{bands}
\end{figure} 
\begin{align}
\mathcal{H}_{FM}(\bold k)=\left(
\begin{array}{cc}
\mathscr A_1(\bo)& \mathscr B(\bo)\\
\mathscr B^\dagger(\bo)& \mathscr A_2(\bo)
\end{array}
\right),
\label{honn}
\end{align}
where
\begin{align}
\mathscr A_1(\bold k)=\left(
\begin{array}{cc}\epsilon_{A1}+m_{\bo}&-v_sf_{\bo}\\
-v_sf_{\bo}^*&\epsilon_{B1} -m_{\bo}\\
\end{array}
\right),
\label{a1}
\end{align}

\begin{align}
\mathscr A_2(\bold k)=\left(
\begin{array}{cc}\epsilon_{A2}+m_{\bo}&-v_sf_{\bo}\\
-v_sf_{\bo}^*&\epsilon_{B2} -m_{\bo}\\
\end{array}
\right),
\label{a2}
\end{align}
\begin{align}
\mathscr B(\bold k)=\left(
\begin{array}{cc}
-v_2f_{\bo}&-v_1f_{\bo} \\
-v_0&-v_2f_{\bo}\\
\end{array}
\right),
\label{bb}
\end{align}
where $\epsilon_{A1}=\epsilon_{B2}=v_s^\prime+z(v_1+v_2)$, and $\epsilon_{B1}=\epsilon_{A2}=v_s^\prime+v_0+zv_2$. The lattice factor is 
 $f_{\bo}= e^{ik_ya/2}\lb 2\cos\sqrt{3}k_xa/2+e^{-3ik_ya/2}\rb,$
and the mass is $m_\bo= 4v_D \sin \frac{\sqrt{3}k_x}{2}\lb \cos \frac{\sqrt{3}k_x}{2}-\cos \frac{{3}k_y}{2}\rb$, with $m_\bo=-m_{-\bo}$.  Notice that  the shifting parameters $\epsilon_{A1} $ and $\epsilon_{B1}$ are the major differences between the bilayer magnon insulator  and that of spin orbit coupled AB-stacked bilayer graphene \cite{mak55}. As a result for  $v_1=v_2=0$, the eigenvalues of the bilayer magnon insulator yields
\begin{align}
&\epsilon_{\pm}^{(1)}=v_s^\prime \pm \sqrt{m_\bo^2 +|v_sf_\bo|^2},\\
&\epsilon_{\pm}^{(2)}=v_s^\prime+v_0 \pm \sqrt{m_\bo^2 +v_0^2 +|v_sf_\bo|^2}.
\label{eig}
\end{align}
At the Dirac points ${\bf K}_\pm= (\pm 4\pi/3\sqrt{3}a, 0)$,  $f_\bo=0$ and $m_\bo=m=3\sqrt{3}v_D$. The lowest band $\epsilon_{-}^{(1)}$ has a Goldstone mode at ${\bf \Gamma}=(0,0)$ when $h=0$.  The evolution of the magnon bands  are shown in  Fig.~\ref{bands}.  For $v_D=v_{1,2}=0$ there are three gapless bands at  ${\bf K}_\pm$ as opposed to AB-stacked bilayer graphene \cite{mcc, mcc1,mcc2,mcc3}.
    
\subsection{Antiferromagnetically coupled layers}

The magnon insulator can also be coupled antiferromagnetically. In this case, the spins on the upper layer are designed to point in opposite direction to those on the lower layer, hence the interlayer couplings become antiferromagnetic with $J_\alpha<0$. To capture the correct magnetic excitations,   we perform  a $\pi$-rotation about  the $S_x$-axis on the upper layer, that is $S_{i,j}^x\to S_{i,j}^x$, $S_{i,j}^y\to -S_{i,j}^y$, $S_{i,j}^z\to -S_{i,j}^z$ (see Appendix~\ref{HHPP}). This rotation does not change the ferromagnetic nature of the upper layer $H_{FM,T}$, but it rotates the spins along the new $z$-axis and changes  the sign of upper-layer out-of-plane DM interaction, $H_{DMI,T}$ and the upper-layer external magnetic field $H_{ext, T}$.   In the HP bosonic representation, the interlayer coupling has the form 
\begin{align}
H_{int.}^{AFM}= \sum_{i\in T, j\in B; \alpha}|v_{\alpha}|[( b_{i}^\dagger b_{i}+ b_{j}^\dagger b_{j}) +( b_{i}^\dagger b_{j}^\dagger+  b_{j} b_{i})].
\end{align}
Introducing the Nambu operators $\Psi^\dagger_{\bold k}= (\psi^\dagger_{\bold k},\psi_{-\bold k})$, the momentum space of the total Hamiltonian can be written as $ H=\frac{1}{2}\sum_{\bold k}\Psi^\dagger_{\bold k}\cdot \mathcal{H}_{AFM}(\bold k)\cdot\Psi_{\bold k}+\text{const.}\label{ham1},$ where $\mathcal{H}_{AFM}(\bold k)$ is a $2N \times 2N$ ( $N$ is the number of sublattices) matrix given by

\begin{align}
\mathcal{H}_{AFM}(\bold k)=\left(
\begin{array}{cc}
\mathcal A(\bo)& \mathcal B(\bo)\\
\mathcal B^*(-\bo)& \mathcal A^*(-\bo)
\end{array}
\right).
\label{eqn9}
\end{align}
The matrices $\mathcal A(\bo)$ and $\mathcal B(\bo)$ are given by

\begin{align}
\mathcal{A}(\bold k)=
\begin{pmatrix}
 \mathscr A_1^\prime(\bo)& 0\\
 0&\mathscr A_2^{\prime *}(-\bo)
\end{pmatrix}.
\label{honn1}
\end{align}

\begin{align}
\mathcal{B}(\bold k)=
\begin{pmatrix}
0 &\mathscr B^\prime(\bo)\\
\mathscr B^{\dagger\prime}(\bo)& 0\\
\end{pmatrix}.
\label{honn1}
\end{align}
where $ \mathscr A_1^\prime(\bo)= \mathscr A_1(\bo)$ and $ \mathscr A_2^\prime(\bo)= \mathscr A_2(\bo)$  with  $\epsilon_{A1}=zv_s+h+z(|v_1|+|v_2|)$, $\epsilon_{B1}=zv_s+h+|v_0|+z|v_2|$,  $\epsilon_{A2}=zv_s-h+|v_0|+z|v_2|$, $\epsilon_{B2}=zv_s-h+z(|v_1|+|v_2|)$. Also $\mathscr B^\prime(\bo)=\mathscr B(\bo)$ with $-v_\alpha\to |v_\alpha|$.
Due to the mass term  $\mathcal A(\bold k)\neq \mathcal A^*(-\bold k)$,  but $\mathcal B(\bold k)=\mathcal B^*(-\bold k)$.

The Hamiltonian is hermitian but not diagonal. It can be diagonalized by the generalized Bogoliubov transformation 
$\Psi_\bo= \mathcal{P}_\bo Q_\bo$, 
where $\mathcal{P}_\bo$ is a $2N\times 2N$ matrix and  $Q^\dg_\bo= (\mathcal{Q}_\bo^\dg,\thinspace \mathcal{Q}_{-\bo})$ with $ \mathcal{Q}_\bo^\dg=(\alpha_{\bo A1}^{\dg}\thinspace \alpha_{\bo B1}^{\dg}\thinspace \alpha_{\bo A2}^{\dg}\thinspace \alpha_{\bo B2}^{\dg})$ being the quasiparticle operators. The matrix $\mathcal{P}_\bo$ satisfies the relations,
\begin{align}
&\mathcal{P}_\bo^\dg \mathcal{H}_{AFM}(\bo) \mathcal{P}_\bo= \epsilon(\bo); \quad \mathcal{P}_\bo^\dg \eta \mathcal{P}_\bo= \eta,
\label{eqna}
\end{align}
\begin{figure}[ht]
\centering
\includegraphics[width=3in]{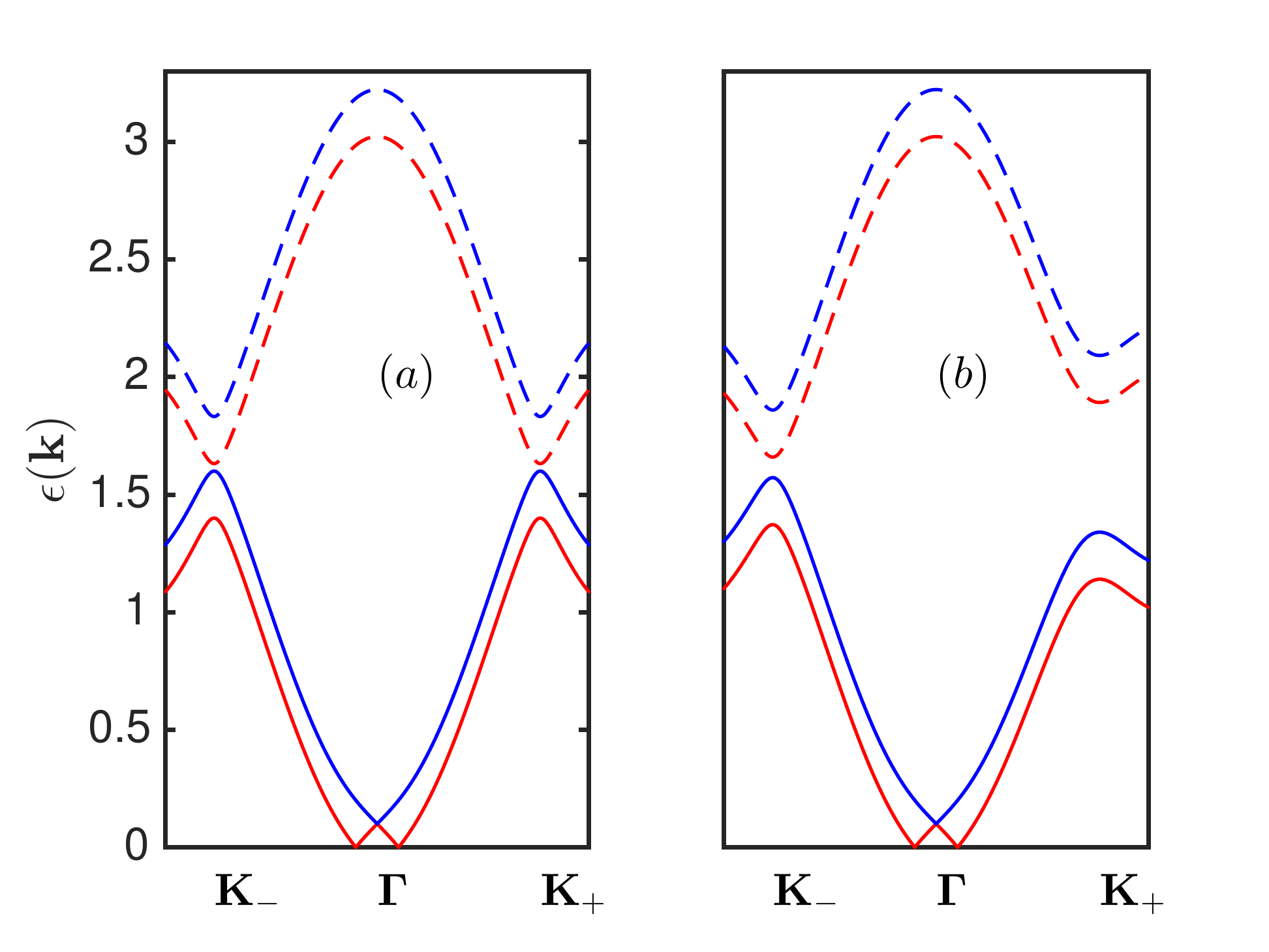}
\caption{Color online.   Magnon bulk bands of  spin-$1/2$ antiferromagnetically coupled  layers along $k_y=0$. The parameters are the same as Fig.~\ref{bands} respectively.}
\label{bandsA}
\end{figure}
with
$\eta=
\text{diag}(
 \mathbf{I}_{N\times N}, -\mathbf{I}_{N\times N} )$, and $\epsilon(\bo)$ is the diagonal matrix of the quasiparticle energy eigenvalues.
Using the fact that $\mathcal{P}_\bo^\dg=\eta \mathcal{P}_\bo^{-1}\eta$, then Eq.~\ref{eqna} is equivalent to saying that we need to diagonalize a non-hermitian Hamiltonian 
 \begin{align}
\eta\mathcal{H}_{AFM}(\bo)=\left(
\begin{array}{cc}
\mathcal A(\bo)& \mathcal B(\bo)\\
-\mathcal B^*(-\bo)& -\mathcal A^*(-\bo)
\end{array}
\right),
\label{non}
\end{align}
 whose eigenvalues are given by $\eta\epsilon(\bo)=[\epsilon_\mu(\bo), -\epsilon_\mu(\bo)]$ and the columns of $\mathcal P_\bo$ are the corresponding eigenvectors.  We could not find an analytical expression in this case.  Figure \ref{bandsA}  shows the positive energy bands.   At the Dirac points $\bold{K}_\pm$, the eigenvalues can be found analytically. They are given by $\epsilon_1=m+\epsilon_{A1}$, $\epsilon_2=m+\epsilon_{B2}$, $\epsilon_{3,4}= \pm h-m+\sqrt{[(\epsilon_{B1}+\epsilon_{A2})/2]^2-v_0^2}$. A careful calculation also shows that antiferromagnetically coupled bilayer ferromagnets behaves very similar to antiferromagnetically coupled bilayer antiferromagnets with N\'eel states, i.e., $J,J_\alpha<0$. For the most part of this paper, we will focus on the special limit $v_1=v_2=0$ for simplicity. In many cases of physical interest, the magnetic field can induce canting of spins in antiferromagnetic systems. This scenario is analyzed in Appendix B.

\section{Topological magnon transports}
\subsection{Magnon edge states}

\begin{figure}[ht]
\centering
  \subfigure[\label{berry_F}]{\includegraphics[width=.45\linewidth]{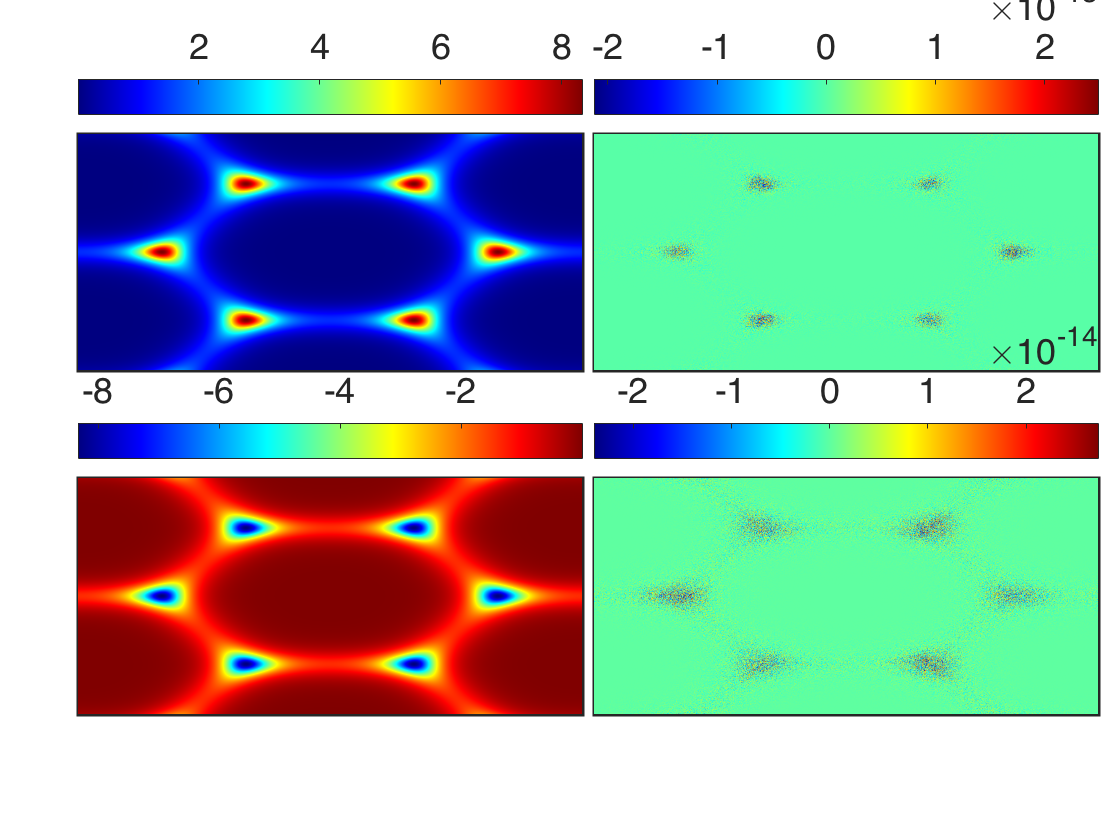}}
   \subfigure[\label{berry}]{\includegraphics[width=.45\linewidth]{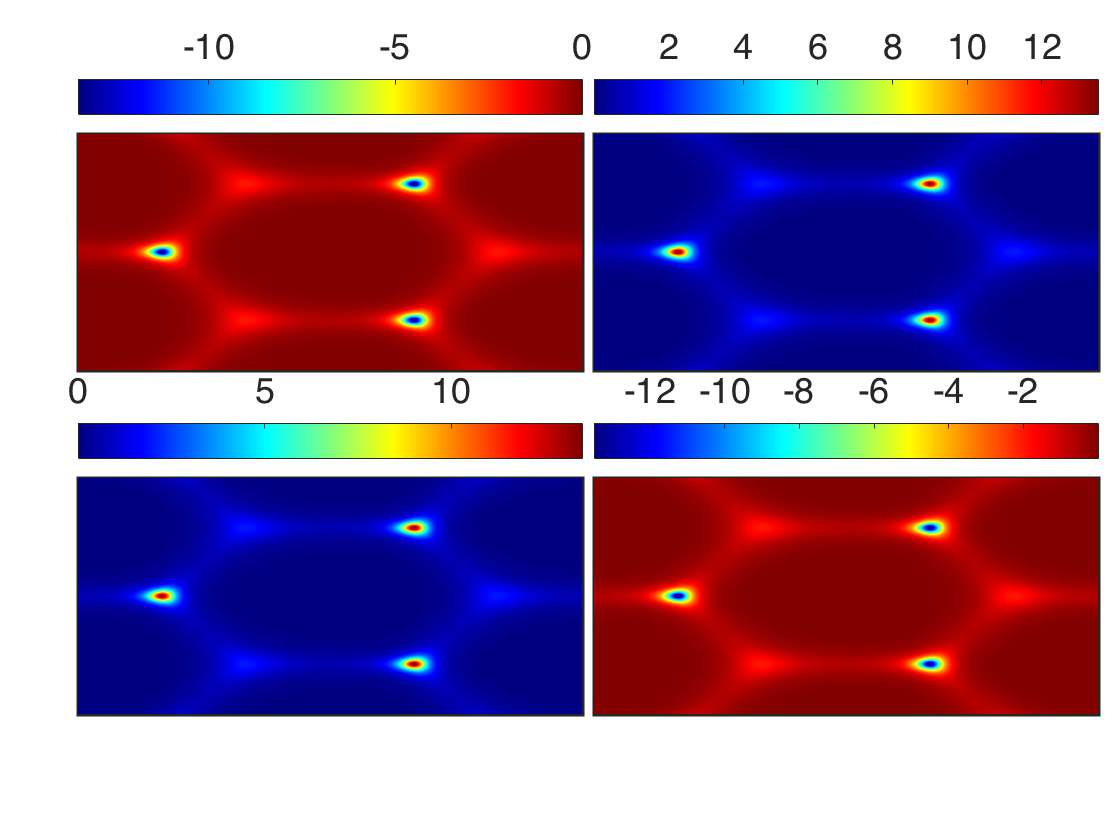}}
\caption{Color online.   Berry curvatures of spin-$1/2$ bilayer honeycomb quantum magnets for bands $\epsilon_{-}^{(1)}$ and $\epsilon_{-}^{(2)}$ (upper panel) and $\epsilon_{+}^{(1)}$ and $\epsilon_{+}^{(2)}$ (lower panel). $(a)$ Ferromagnetic coupling.  $(b)$ Antiferromagnetic coupling.  The parameters are the same as Figs.~\ref{bands}(b) and \ref{bandsA}(b) respectively. }
\label{berry_A}
\end{figure}

\begin{figure*}[!]
\centering
  \subfigure[\label{edge}]{\includegraphics[width=.45\linewidth]{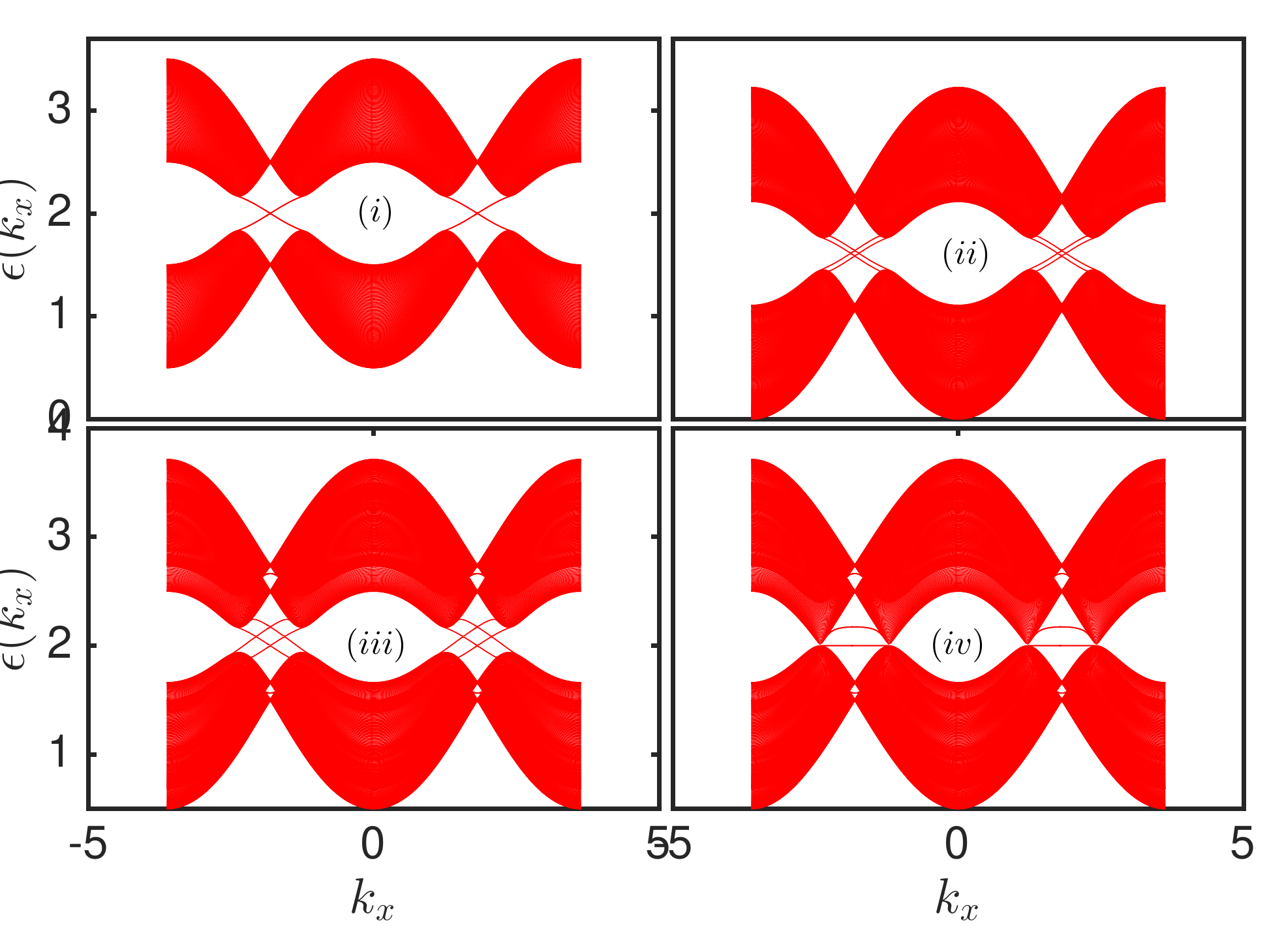}}
   \quad
   \subfigure[\label{edgeA}]{\includegraphics[width=.45\linewidth]{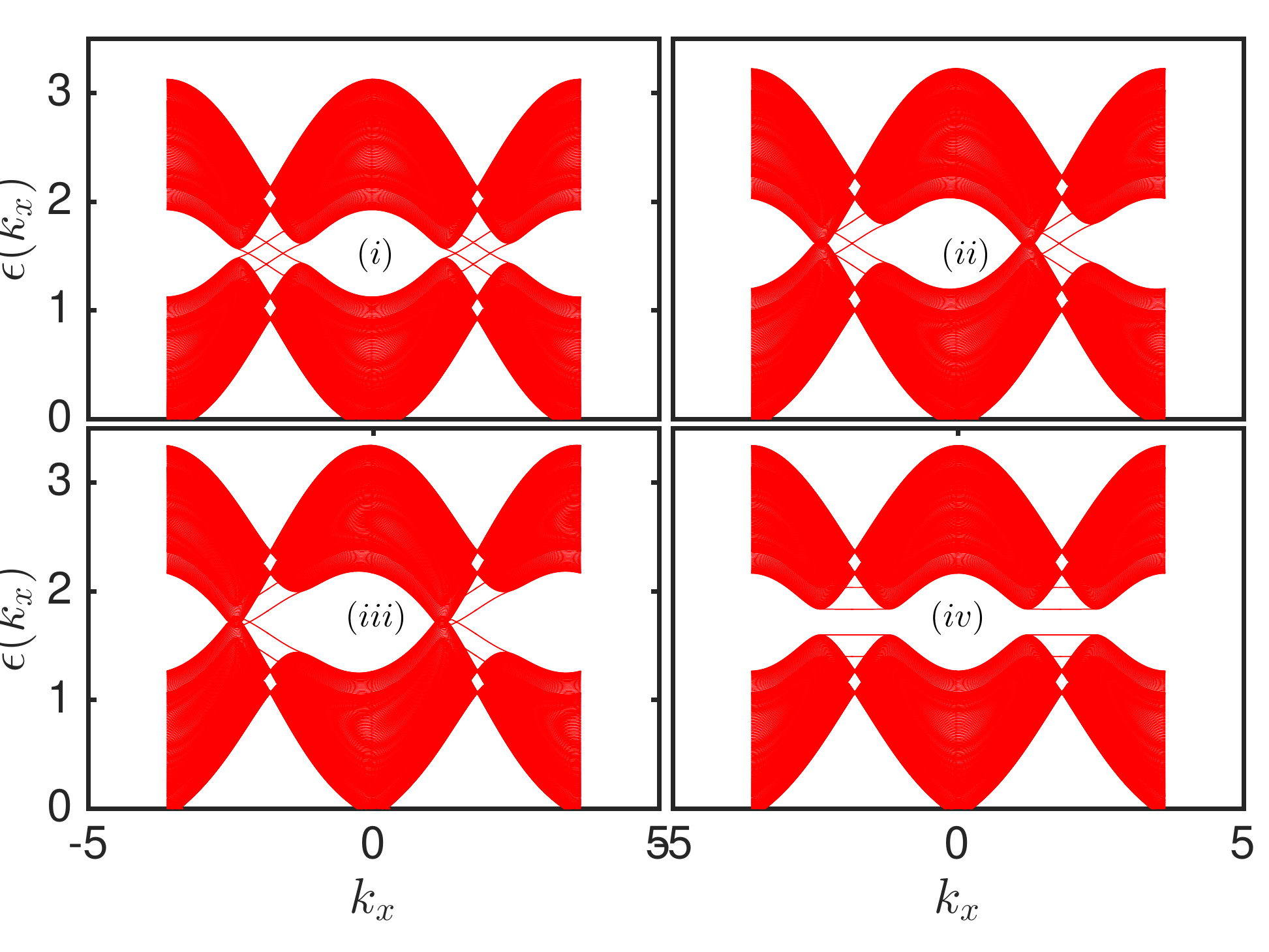}}
\caption{Color online. Magnon edge states of bilayer spin-$1/2$ honeycomb ferromagnet in the zigzag geometry for $v_1=v_2=0$ and $v_s=0.5,~h=0.1$. $(a)$ Ferromagnetic coupling, $(i)~v_D=0.1,~v_0=0;~(ii)~v_D=0.1, v_0=0.05; ~(iii)~v_D=0.1, v_0=0.25; ~(iv)~v_D=0.0, v_0=0.5$. $(b)$ Antiferromagnetic coupling, $(i)~v_D=0.1, v_0=0.05; ~(ii)~v_D=0.1, v_0=0.25;~(iii)~v_D=0.1, v_0=0.5; ~(iv)~v_D=0.0, v_0=0.5$. The bands in Figs.~\ref{bands}(b) and \ref{bandsA}(b) correspond to $a~(iii)$ and $b~(ii)$. }
   \label{Edge}
\end{figure*}

\begin{figure}[ht]
\centering
  \subfigure[\label{FMedge}]{\includegraphics[width=.45\linewidth]{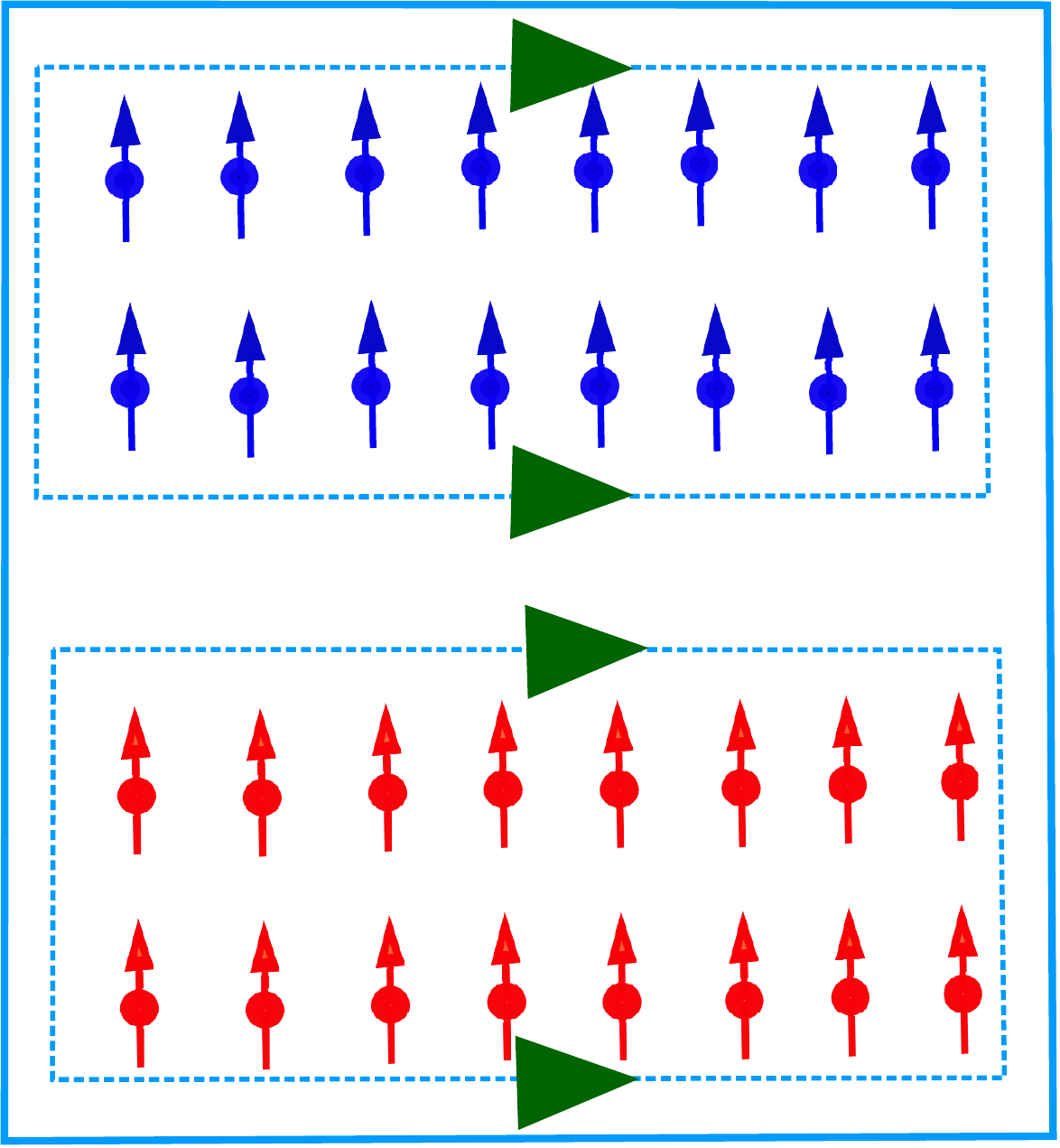}}
   \quad
   \subfigure[\label{AFMedge}]{\includegraphics[width=.45\linewidth]{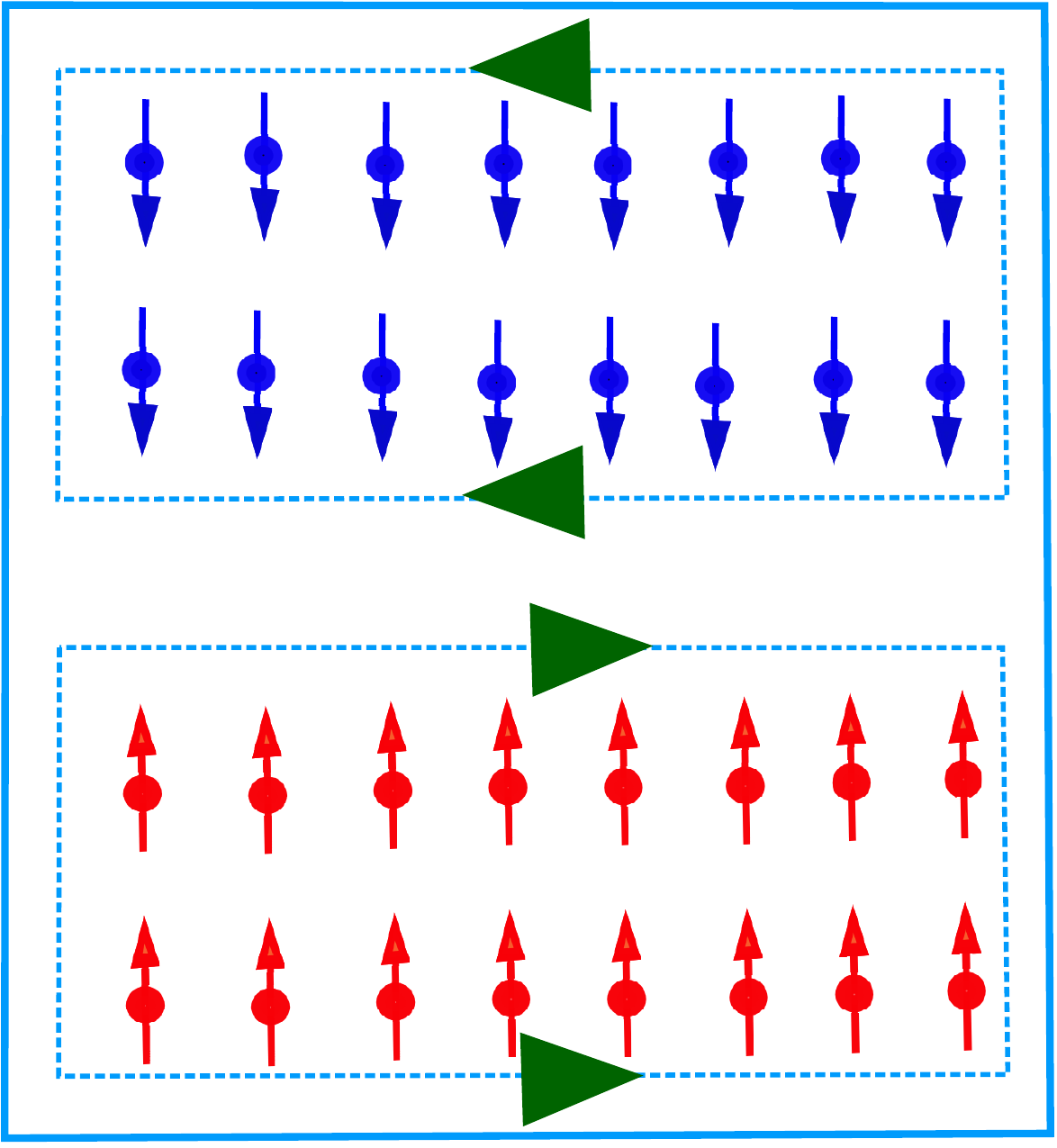}}
\caption{Color online. Schematics of the propagation of chiral magnon edge states.  $(a)$  Ferromagnetically coupled layers. $(b)$ Antiferromagnetically coupled layers. }
\end{figure}

We now study the main purpose of this paper --- thermal Hall transports of magnons. As the DMI plays the same role as spin-orbit coupling \cite{mak55}, topological effects in magnon insulators stem from  nontrivial properties of the chirality-induced Berry curvatures. This directly implies the propagation of magnon edge states on the edges of each layer. Having diagonalized the Hamiltonians, the Berry curvature of the magnon bulk  bands can be written  as
\begin{eqnarray}
\Omega_{ij;\mu}(\bold k)=-2\sum_{\mu^\prime\neq \mu}\frac{\text{Im}[ \braket{\mathcal{P}_{\bo\mu}|v_i|\mathcal{P}_{\bo\mu^\prime}}\braket{\mathcal{P}_{\bo\mu^\prime}|v_j|\mathcal{P}_{\bo\mu}}]}{\lb\epsilon_{\bo\mu}-\epsilon_{\bo\mu^\prime}\rb^2},
\label{chern2}
\end{eqnarray}
where $v_{i}=\partial \mathcal{H}_{FM}(\bold k)/\partial k_{i}$ defines the velocity operators with $i,j=x,y$. The columns of $\mathcal{P}_{\bo\mu}$ are the eigenvectors, and $\mu$ labels the bands. For antiferromagnetically coupled layers,  $v_{i}=\partial [\eta\mathcal{H}_{AFM}(\bold k)]/\partial k_{i}$ and  $\mathcal{P}_{\bo\mu}$ is paraunitary. In the antiferromagnetic case the Berry curvature can be written alternatively as
\begin{align}
 \Omega_{ij;\mu}(\bo)=-2\text{Im}[\eta\mathcal (\partial_{k_i}\mathcal P_{\bo\mu}^\dg)\eta(\partial_{k_j}\mathcal P_{\bo\mu})]_{\mu\mu},
 \label{bc1}
 \end{align}
which can be used for systems with explicit analytical form of $\mathcal P_{\bo\mu}$.
The Berry curvature  for ferromagneticcally coupled layers is  shown in  Fig.~\ref{berry_F} and antiferromagneticcally coupled layers in Fig.~\ref{berry}.  In the former,  only the Berry curvatures corresponding to the bands $\epsilon_\pm^{(1)}$ are peaked at the Dirac points. This is due to the biased nature of the system.  On the other hands, the latter  Berry curvatures for each band is suppressed at three corners of the Brillouin zone [see the corresponding bands in Fig.~\ref{bandsA}(b)]. 
Although magnon Hall transports are induced by the Berry curvatures, the Chern numbers can still be defined for bosonic systems as the integration of the Berry curvature over the first Brillouin zone given by,
 \begin{equation}
\mathcal{C}_\mu= \frac{1}{2\pi}\int_{{BZ}} dk_xdk_y~ \Omega_{xy; \mu}(\bold k).
\label{chenn}
\end{equation}  
For ferromagnetically coupled layers, the Chern number of the bands $\epsilon_\pm^{(1)}$ and $\epsilon_\pm^{(2)}$  are given by $\mathcal{C}_\mu=[2,-2,0,0]$ and antiferromagnetically coupled layers, we find $\mathcal{C}_\mu=[2,-2,2,-2]$. We see that the sum of the Chern numbers is zero as can also be seen from the Berry curvatures.  Similar to AB-stacked bilayer graphene \cite{mak55}, the Chern numbers are double that of single layer \cite{sol}.   This implies that the edge states are also double that of single layer [Fig.~\ref{edge} $(i)$]  as depicted in Figs.~\ref{edge}  $(ii)$ and $(iii)$ and \ref{edgeA}  $(i)-(iii)$.  We also observe magnon edge states at zero DMI [Figs.~\ref{edge} and \ref{edgeA} $(iv)$], which is not protected by any Chern number  similar to AB-stacked bilayer graphene \cite{castt}. The magnon edge states for ferromagnetically and antiferromagnetically coupled layers differ by the direction of propagation  as shown schematically in Figs.~\ref{FMedge} and \ref{AFMedge}.   

\subsection{Magnon Hall and Spin Nernst effects}
\begin{figure}[!]
\centering
  \subfigure[\label{mags}]{\includegraphics[width=.45\linewidth]{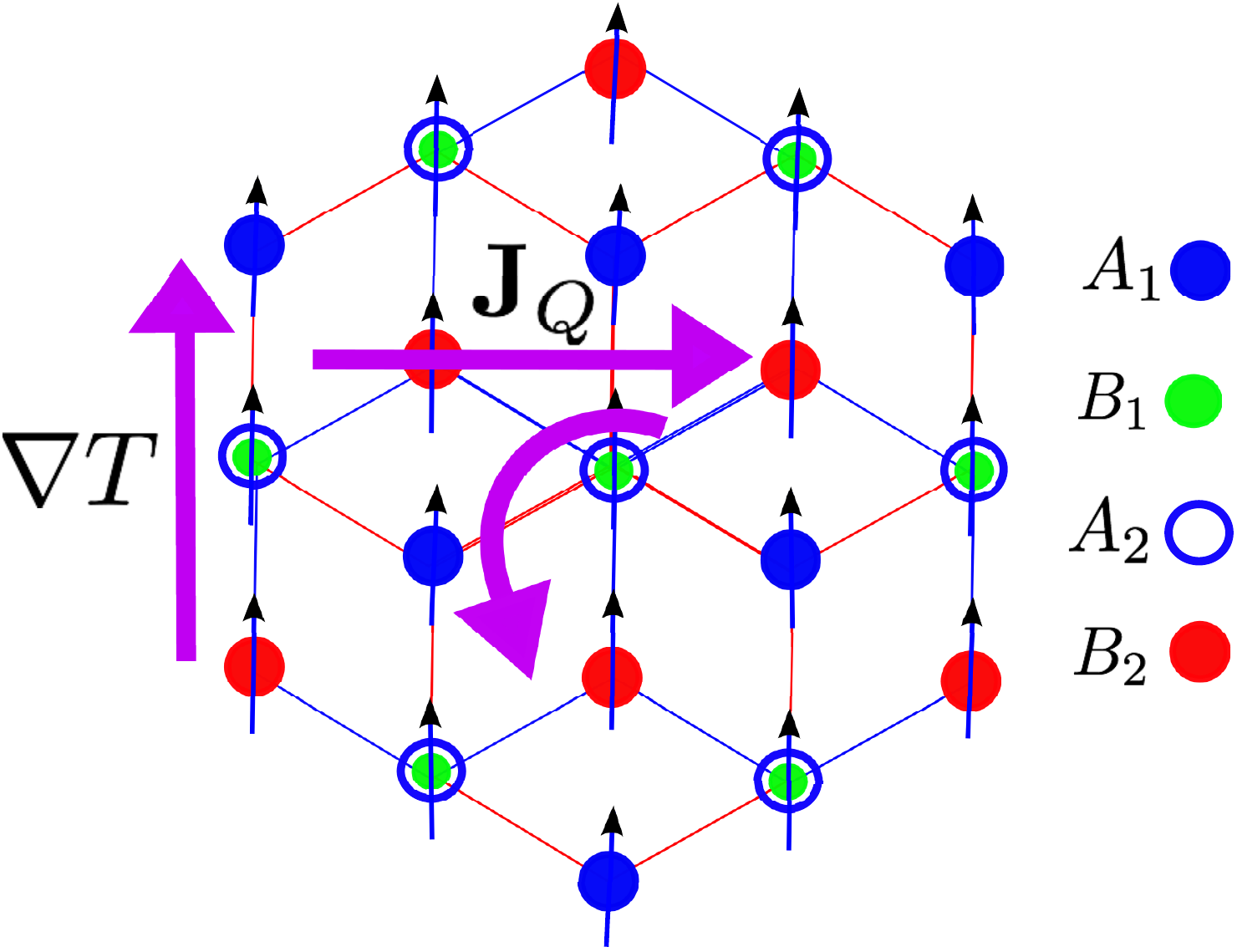}}
   \quad
   \subfigure[\label{magt}]{\includegraphics[width=.45\linewidth]{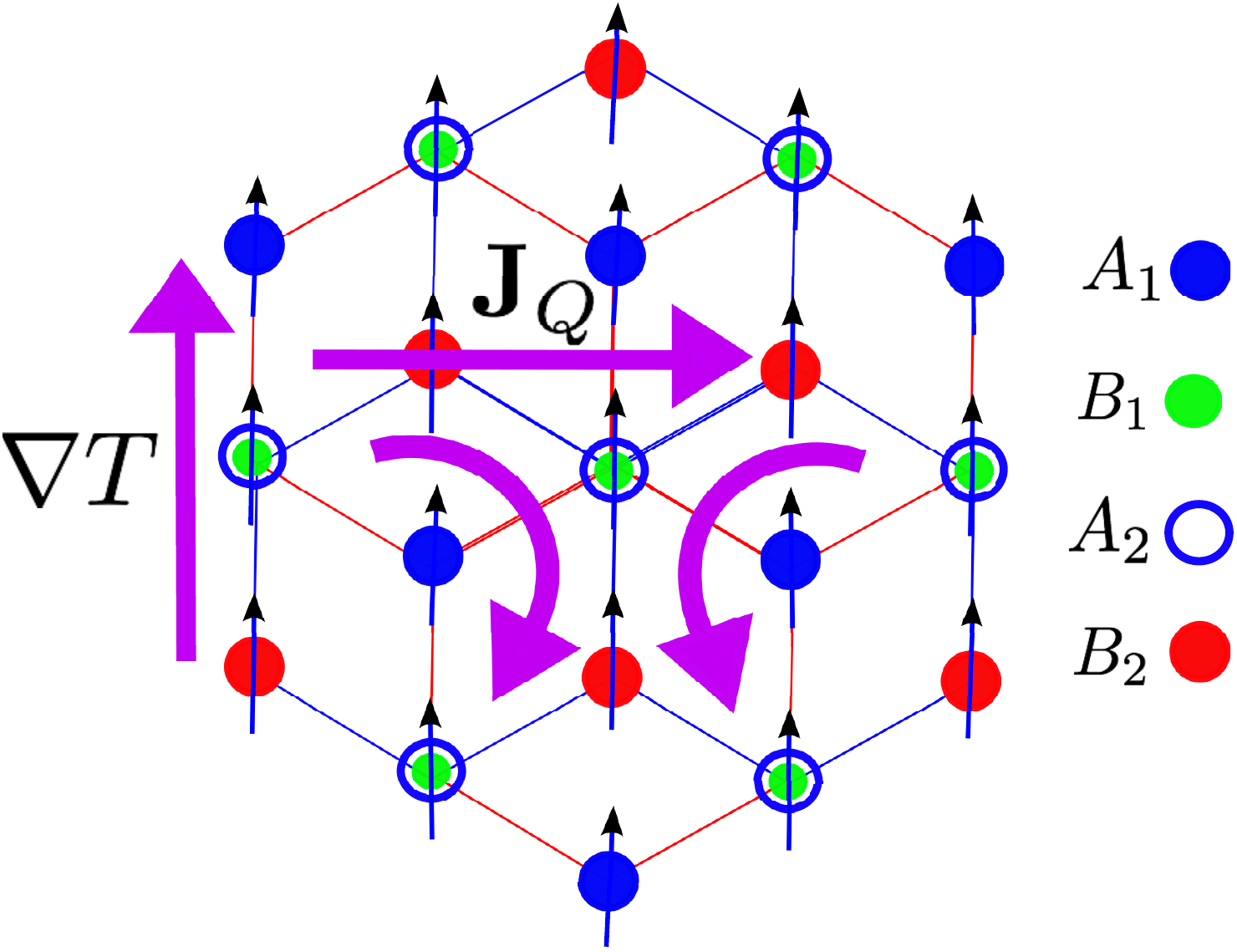}}
\caption{Color online. Schematics of magnon Hall effect $(a)$ and spin Nernst effect $(b)$  in ferromagnetically coupled layers.  In magnon Hall effect, the propagation of magnon is deflected  by the DMI upon the application of temperature gradient $\nabla T$ and an induced heat current ${\bf J}_Q$; whereas for spin Nernst effect, opposite spins propagate in different directions. Notice that for antiferromagnetically coupled layers, the schematic representation of magnon Hall effect is equivalent to $(b)$, since the upper layer has an opposite DMI as shown in Eq.~\ref{eqn9}.  }
\end{figure}

Theoretically, thermal Hall effect of magnons is manifested due to the nontrivial topology of the magnon band structures  induced by the DMI. The non-vanishing Berry curvatures induce an effective magnetic field in  the system, upon the application of a longitudinal temperature gradient $-\boldsymbol \nabla T$.   Figure~\ref{mags} shows the schematic representation of magnon Hall effect, in which a longitudinal temperature gradient induces a transverse heat current ${\bf J}_Q$. The propagation of magnon in the bilayer system is deflected by the DMI on the top and bottom layers.  This  leads to thermal Hall effect \cite{alex0} characterized  by the transverse thermal Hall conductivity   given by \cite{alex2,shin1} 
$
\kappa_{xy}=-\frac{k_B^2 T}{\hbar V}\sum_{\bo}\sum_{\mu=1}^N c_2[ g\lb\epsilon_{\bo \mu}\rb]\Omega_{xy; \mu}(\bold k),$ where $V$ is the volume of the system, $k_B$ is the Boltzmann constant, $T$ is the temperature, $g(\epsilon_{\bo\mu})=[e^{{\epsilon_{\bo\mu}}/k_BT}-1]^{-1}$ is the Bose function,  $c_2(x)=(1+x)\lb \ln \frac{1+x}{x}\rb^2-(\ln x)^2-2\text{Li}_2(-x),$ and $\text{Li}_2(x)$ is a dilogarithm. 
\begin{figure}[!]
\centering
  \subfigure[\label{th_f}]{\includegraphics[width=.45\linewidth]{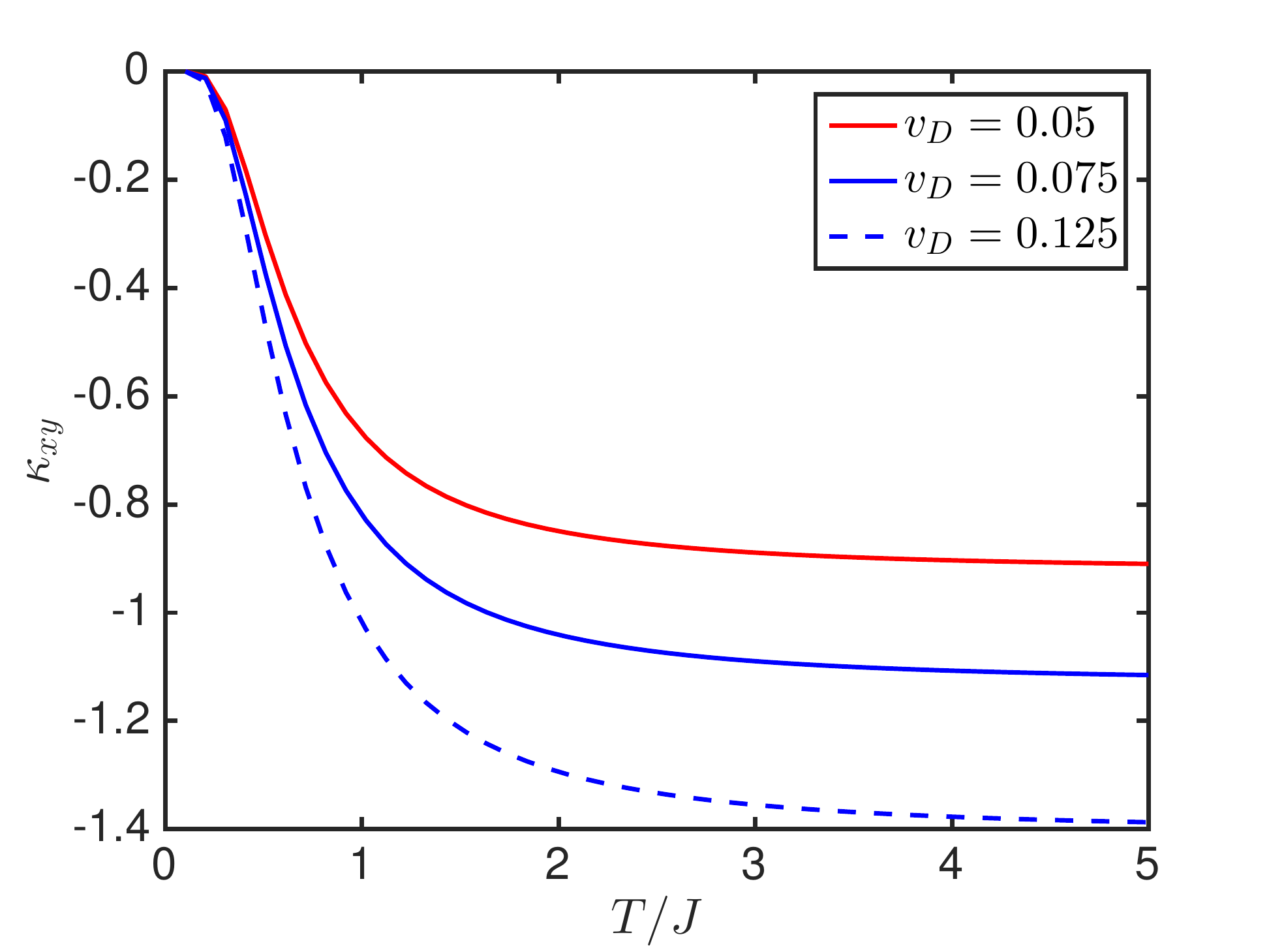}}
   \quad
   \subfigure[\label{th1_f}]{\includegraphics[width=.45\linewidth]{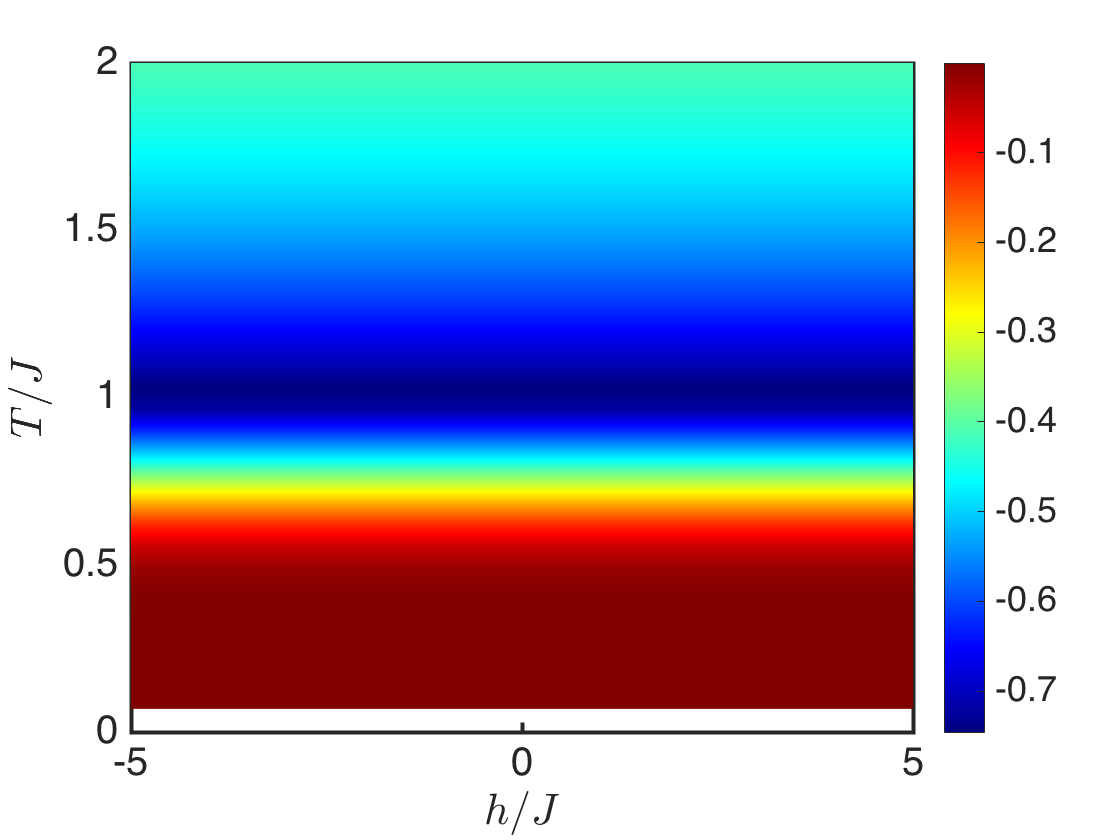}}
   \quad
   \subfigure[\label{sn_f}]{\includegraphics[width=.45\linewidth]{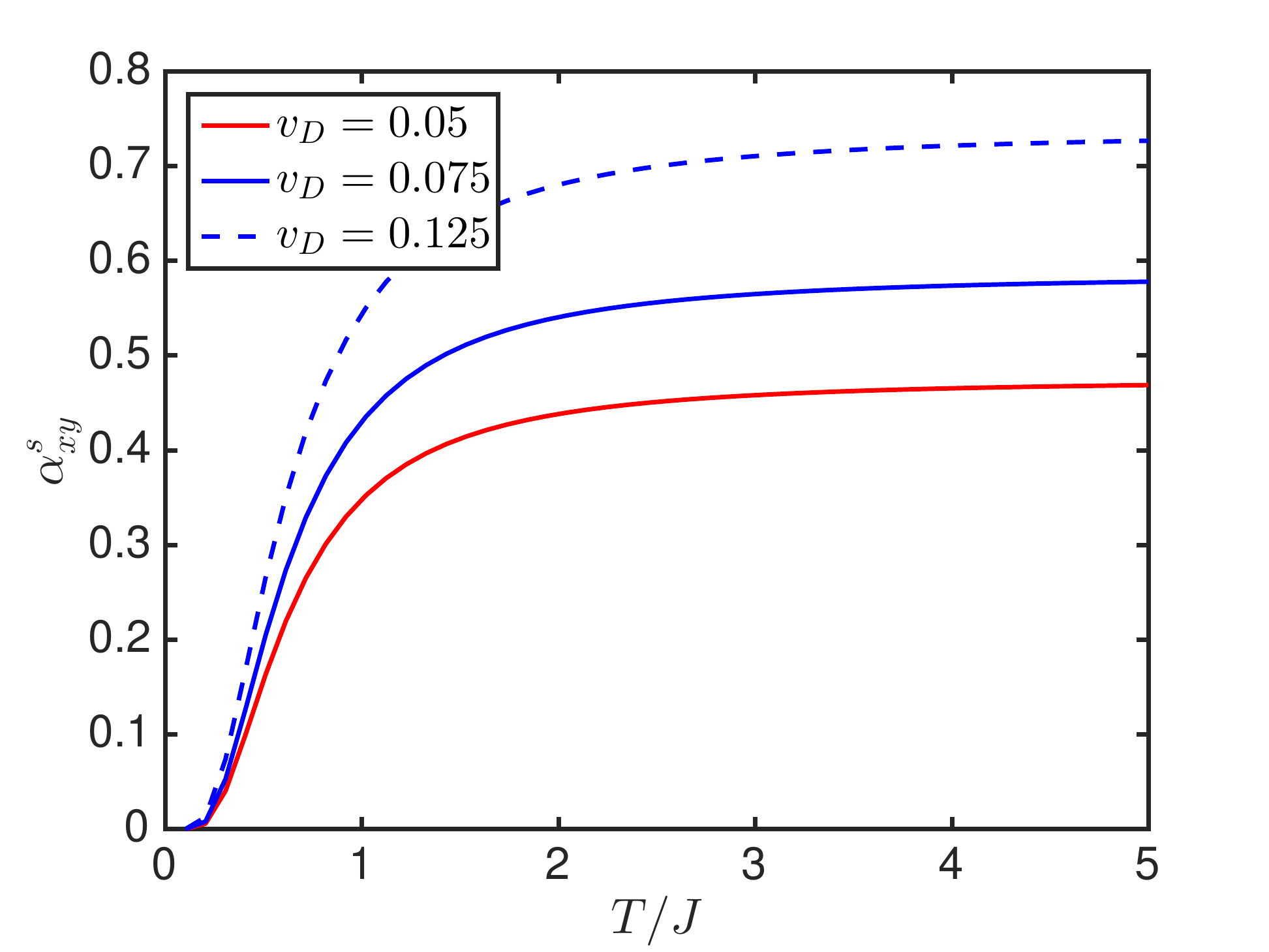}}
   \quad
   \subfigure[\label{sn1_f}]{\includegraphics[width=.45\linewidth]{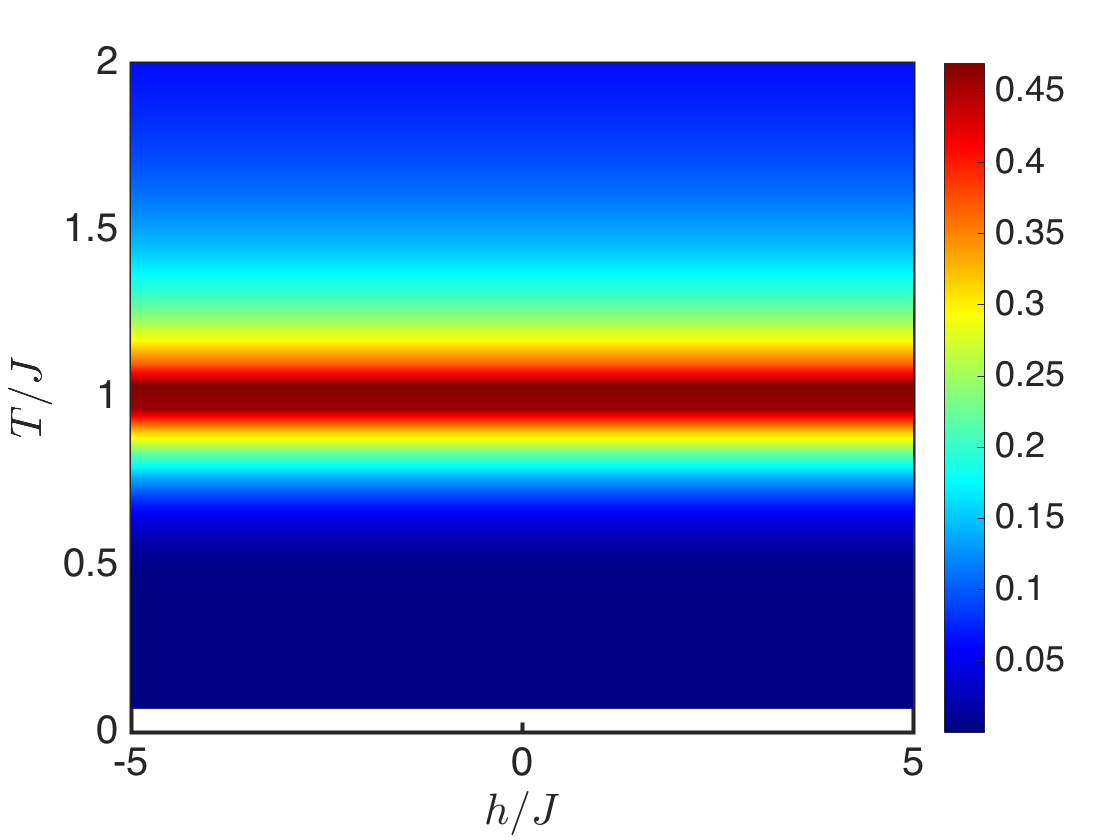}}
\caption{Color online. Plot of $\kappa_{xy}$ and $\alpha_{xy}^s$ for spin-$1/2$ ferromagnetically coupled layers. $(a)$ $\kappa_{xy}$ vs.  $T/J$, $(b)$ Contour plot of $\kappa_{xy}$ in $T/J$ and $h/J$ plane;  $(c)$ $\alpha_{xy}^s$ vs.  $T/J$;   $(d)$ Contour plot of $\alpha_{xy}^s$ in $T/J$ and $h/J$ plane. The parameters  for $(a)$ and $(c)$ are $k_B=\hbar=1$,  $v_s=v_0=0.5$, $h=\pm 0.5$, $v_1=v_2=0$, and  several values of $v_D$. For $(b)$ and $(d)$,  $v_s=v_0=0.5; v_D=0.05$. }
\end{figure}
\begin{figure}
\centering
  \subfigure[\label{th_a}]{\includegraphics[width=.45\linewidth]{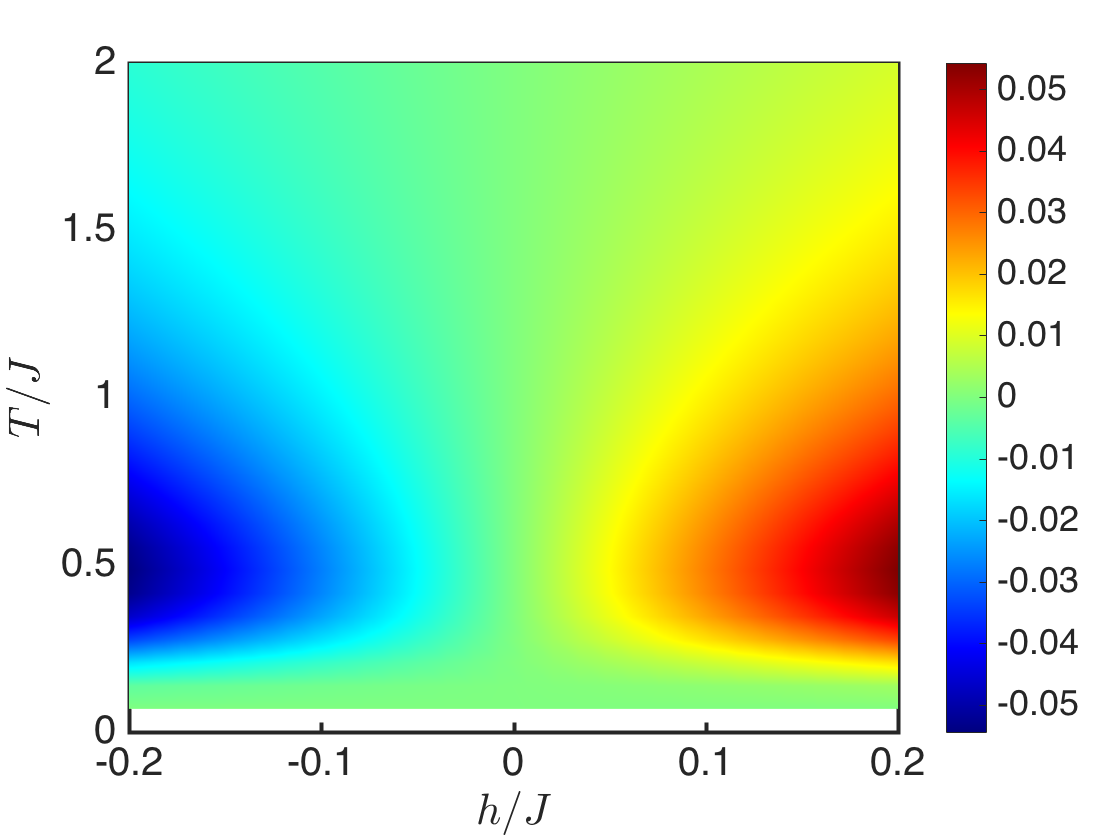}}
   \quad
   \subfigure[\label{sn_a}]{\includegraphics[width=.45\linewidth]{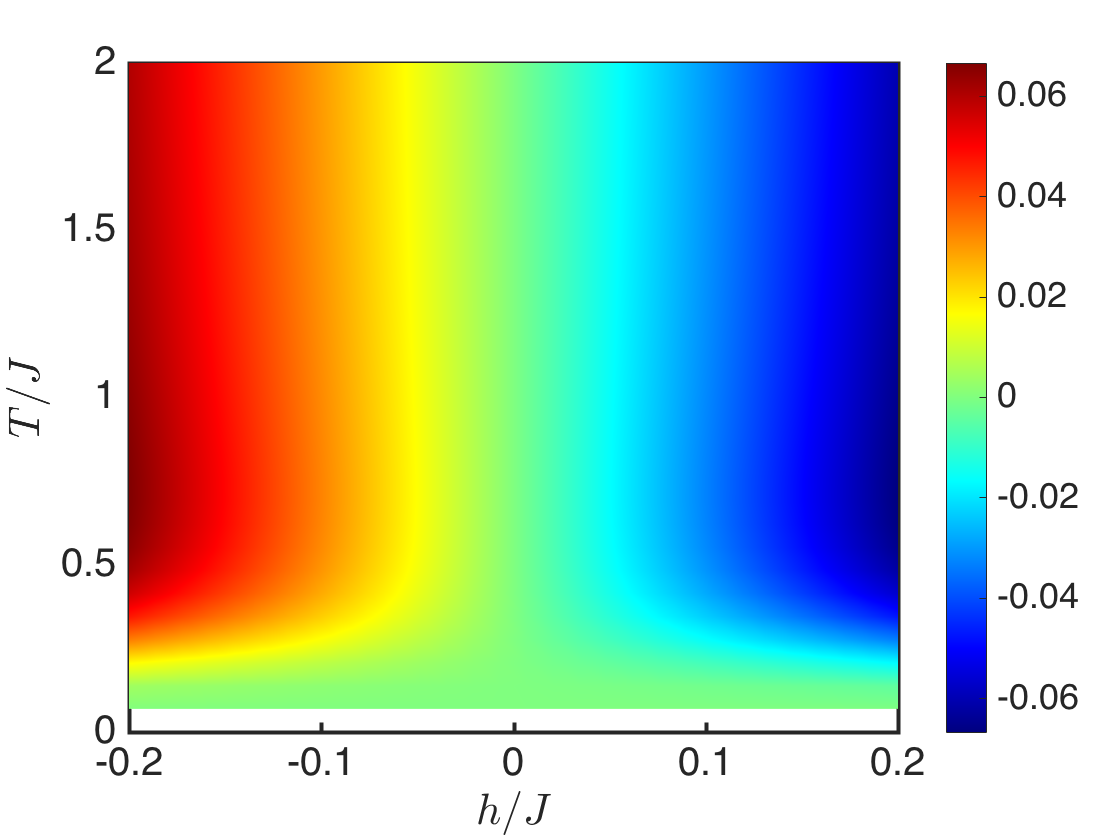}}
\caption{Color online. $(a)$  Contour plot of $\kappa_{xy}$ in $T/J$ and $h/J$ plane for spin-$1/2$ antiferromagnetically coupled layers. $(b)$ Contour plot of $\alpha_{xy}^s$ in $T/J$ and $h/J$ plane. The parameters are  $k_B=J=\hbar=1$,  $v_s=0.5~v_0=0.5,v_D=0.05;~v_1=v_2=0$. }
\end{figure}

In what follows, we focus on the simplified model $v_1=v_2=0$  with nonzero $v_0,v_D$. The trend of the thermal conductivity is shown in Fig.~\ref{th_f} for several values of $v_D$ and fixed $v_0=0.5$.  At  $T=0$,  $\kappa_{xy}=0$ and decreases exponentially for large temperature approaching a constant value for very large temperature but never changes sign. Increasing the DMI, $v_D$, decreases $\kappa_{xy}$. The contour plot of $\kappa_{xy}$ in Fig.~\ref{th1_f} shows no sign change by reversing the magnetic field. 
As previously mentioned, magnon edge state propagation also carries a transverse spin current which leads to spin Nernst effect depicted in Fig.~\ref{magt}. It can be understood as two copies of magnon Hall effects with opposite spins. It is  characterized by a conductivity given by \cite{alex7}
$\alpha_{xy}^s={k_B}V^{-1}\sum_{\bo\mu}c_1\lb n_\mu\rb\Omega_{xy; \mu}(\bold k),$
where  $c_1(x)=(1+x)\ln(1+x)-x\ln x$. Due to the Berry curvature,  $\alpha_{xy}^s$ has a similar feature as $\kappa_{xy}$ except that it is now positive as shown in Fig.~\ref{sn_f}. The spin Nernst conductivity also vanishes at zero temperature as there are no thermal excitations but the DMI, $v_D$ increases $\alpha_{xy}^s$. We see that $\alpha_{xy}^s$ does not show any sign change for all parameters considered as shown in Fig.~\ref{sn1_f}.

 For antiferromagnetically coupled layers,  the propagation of magnons are deflected in opposite directions by opposite signs of DMI on each layer (see Eq.~\ref{eqn9}). In addition, the Berry curvature changes sign as the magnetic field is reversed (not shown). Hence, the topology of the system is different from ferromagnetically coupled layers. This is manifested explicitly in the conductivities as shown in Figs.~\ref{th_a} and \ref{sn_a}.  The sign change in $\kappa_{xy}$ and $\alpha_{xy}^s$ as the magnetic field is reversed is the manifestation of the band topology of the system for   antiferromagnetically coupled layers. A different sign change has been studied on the kagome lattice. In this case, Ref.~[\onlinecite{alex44}] explains the sign change in $\kappa^{xy}$  as a consequence of the sign change in  Berry curvature of the highest band and Ref.~[\onlinecite{alex4}] argues that the sign change in $\kappa^{xy}$ is a consequence of the propagation of the magnon edge states, however, with a NNN interaction. The origin of the sign change on the  kagome ferromagnets is still not well-understood theoretically. In the present case with antiferromagnetic coupling, reversing the magnetic field flips the spin moments on both layers and thus changes the sign of the spin and bosonic operators in HP transformation (see Appendix~\ref{HHPP}). This has an important consequence on the sign of the DMI on each layer and affects the signs of the conductivities.

\section{Conclusion} 
We have presented a detail study of magnon Hall transports in AB-stacked bilayer honeycomb magnon insulators. We show that the interlayer couplings between two ferromagnetic magnon insulators can be treated ferromagnetically or antiferromagnetically. For ferromagnetic coupling, we present explicit calculation based on magnon tight binding model obtained from linearized HP transformation.  We show that the system behaves similarly to AB-stacked bilayer graphene and possesses nontrivial topological magnon bulk bands and Berry curvatures. However, for magnon insulators the propagation of edge states give rise to thermal Hall effect and spin Nernst effect. We show that for ferromagnetically coupled layers, the conductivities characterizing these effects are very similar to single-layer honeycomb magnon insulator with no sign change.  For antiferromagnetically coupled layers, we show that the topological  nature of the magnon bands is different.  Magnon edge states propagate in opposite directions and the conductivities of thermal Hall response show a sign change as the magnetic field is reversed.     

To the best of our knowledge, thermal Hall effect of magnons  has been observed experimentally only in ferromagnetic insulators without inversion symmetry.  The Lieb and honeycomb lattices still await experimental observation. In these systems, spin-orbit coupling (DMI) can be induced in many different ways.  For the honeycomb lattice, it is also possible to utilize ultracold bosonic atoms trapped in optical lattice at finite but low temperatures, since the bosonic tight binding model is reminiscent of tight binding model in electronic systems, where the Haldane model has been realized experimentally in optical fermionic lattice \cite{jott}.   The most important result of our study  is that it opens new possibilities to search for bilayer magnon Hall transports  in other lattices without inversion symmetry (such as the kagome and pyrochlore lattices), where nontrivial topological effects and magnon Hall effect have been realized experimentally.  Our study  also generalizes  the study of magnon spintronics  to bilayer systems. This model  can also be studied using the spinon (Schwinger boson) representation. The Hamiltonian in this representation is an 8-band model, which can be regarded as two copies of Holstein-Primakoff Hamiltonians, one for each spin degree of freedom. The magnon bands and the associated magnon edge states are similar to that of gated  AB-stacked bilayer graphene \cite{zhe}. Both representations give similar results, however  recent experimental results on the kagome magnet Cu(1-3, bdc) confirms that the Holstein-Primakoff boson representation gives a better estimate than the spinon   representation \cite{alex6a, alex6}.

\section*{Acknowledgments} Research at Perimeter Institute is supported by the Government of Canada through Industry Canada and by the Province of Ontario through the Ministry of Research
and Innovation.
\appendix

 \section{Holstein-Primakoff representation}
 \label{HHPP}
 This appendix presents the well-known Holstein-Primakoff representation of spin operators. At low temperature, only few magnons are thermally excited. The linearized Holstein-Primakoff (HP) representation can be used to calculate the magnetic spin excitations. For the bilayer system, there are four sublattices, the  HP representation  for ferromagnetically coupled layers are given by
 
\begin{align}
&S_{j \alpha}^{ z}= S-b_{j\alpha}^\dagger b_{j\alpha},\\&
 S_{j \alpha}^{ y}=  i\sqrt{\frac{ S}{2}}(b_{j\alpha}^\dagger -b_{j\alpha}),\\&
 S_{j\alpha}^{ x}=  \sqrt{\frac{ S}{2}}(b_{j\alpha}^\dagger +b_{j\alpha}),
 \end{align}
where $\alpha=A_1,B_1, A_2, B_2$ label the sublattices. Similar representation holds for spins at site $i$. When the magnetic field changes sign as $H_{ext,\tau}=h\sum_j
S_{j,z}^\tau$, we imagine that the spins on both layers point in the direction of the field. The HP transformation is the same, except that we have to take $S_{j \alpha}^{ z}= -S + b_{j\alpha}^\dagger b_{j\alpha}$, which accounts for spins pointing  in the negative $z$-direction. The magnon band structures are the same in both cases.

\begin{figure}[ht]
\centering
\includegraphics[width=3in]{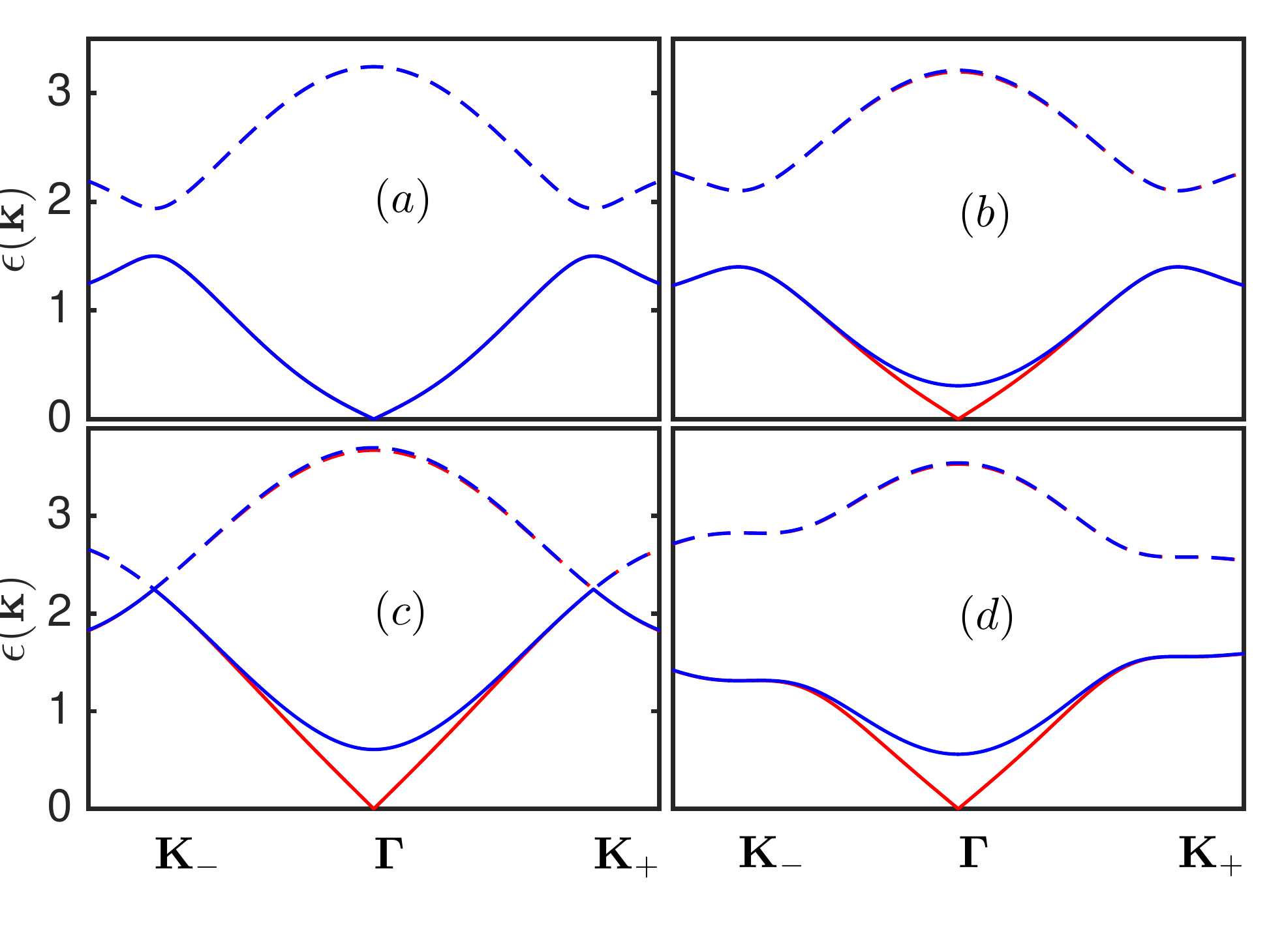}
\caption{Color online.   Magnon bands of spin-$1/2$ antiferromagnetically coupled  layers at zero magnetic field along $k_y=0$ at  $h=0,~v_s=0.5$:  $(a)$ $v_{0}=0.25$, $v_1=v_2=v_D=0$. $(b)$  $v_{1}=0.25$, $v_0=v_2=v_D=0$. $(c)$ $v_{2}=0.25$, $v_0=v_1=v_D=0$. $(d)$ $v_0=v_1=v_2=v_{D}=0.125$..}
\label{bandsAh}
\end{figure}

For antiferromagnetically coupled layers, the spins on the top layer are designed to point in opposite direction to those on the lower layer or vice versa, and the magnetic field is applied along the positive $z$-direction. The interlayer coupling are replaced with $J_\alpha\to -|J_\alpha|$. To capture the correct magnetic excitations, we write two  different HP transformations for the top and bottom layers. For the  bottom layer with up pointing spins we write
\begin{align}
&S_{j \alpha}^{ z}= S-b_{j\alpha}^\dagger b_{j\alpha}\label{ia},\\&
 S_{j \alpha}^{ y}=  i\sqrt{\frac{ S}{2}}(b_{j\alpha}^\dagger -b_{j\alpha})\label{ib},\\&
 S_{j\alpha}^{ x}=  \sqrt{\frac{ S}{2}}(b_{j\alpha}^\dagger +b_{j\alpha})\label{ic},
 \end{align}
where $\alpha=A_1,B_1$ label the sublattices on the bottom layer. For the top layer with down pointing spins we write
\begin{align}
&S_{j \alpha}^{ z}= -S+ b_{j\alpha}^\dagger b_{j\alpha}\label{i},\\&
 S_{j \alpha}^{ y}=  i\sqrt{\frac{ S}{2}}(b_{j\alpha} -b_{j\alpha}^\dagger)\label{ii},\\&
 S_{j\alpha}^{ x}=  \sqrt{\frac{ S}{2}}(b_{j\alpha}^\dagger +b_{j\alpha})\label{iii},
 \end{align}
where $\alpha=A_2,B_2$ label the sublattices on the top layer. Notice the signs of Eq.~\ref{i} and \ref{ii}; see for example  Ref.~[\onlinecite{ll}].  This corresponds exactly to rotating the spins on the upper layer by $\pi$ about the $S_x$ direction to align them in the same direction as the bottom layer. When the magnetic field changes sign, we imagine that the spins on the top and bottom layers change direction as well, then  Eqs.~\ref{ia}--\ref{ic} apply to the top layer and Eqs.~\ref{i}--\ref{iii} apply to the bottom layer.  In this case, the magnon band structures are very similar in both cases, but the topology of the system is different as shown above. At zero magnetic magnetic field, time-reversal symmetry is preserved and the band structure is doubly degenerate for $v_1=v_2=0$ as shown in Fig.~\ref{bandsAh} (a). Notice the linear dispersion near $\bold{\Gamma}$, which is an artifact of the low-energy excitations of antiferromagnetic quantum magnets.  

\section{ Field-induced canting in bilayer antiferromagnets}
 In many cases of physical interest, the magnetic field induces canting of spins in antiferromagnets. In  this section, we consider this scenario in antiferromagnetically coupled layers governed by the Hamiltonian 
 \begin{align}
H&=J\sum_{\la i, j\ra}{\bf S}_{i}^\tau\cdot{\bf S}_{j}^\tau+D\sum_{\la \la i,j\ra\ra}\nu_{ij} \hat{\bf z}\cdot{\bf S}_{i}^\tau\times{\bf S}_{j}^\tau-h\sum_{i}S_{i,z}^\tau,\label{model}\nonumber\\&+J_0\sum_{i\in T; j\in B}{\bf S}_{i}\cdot{\bf S}_{j},
\end{align}
where $J$ is a nearest-neighbour  interaction, ${D}$ is a staggered DMI allowed by the next-nearest-neighbour  triangular plaquettes on the honeycomb lattice, and $\nu_{ij}=\pm 1$.   The Zeeman field is $h$ in units of $g\mu_B$ and $J_0>0$ is the antiferromagnetic interlayer coupling.

For antiferromagnetically coupled ferromagnets considered above, $J<0$. Because of antiferromagnetic coupling between the layers, the magnetic field can introduce canting on each layer. In this case,  we  rotate  the coordinate axes such that the $z$-axis coincides with the local direction of the classical polarization.  The appropriate rotation on the two sublattices is given by 
\begin{align}
&S_{i, \tau}^x=\pm S_{i, \tau}^{\prime x}\sin\chi   \pm S_{i, \tau}^{\prime z}\cos\chi,\label{trans1}\nonumber\\&
S_{i, \tau}^y=\pm S_{i, \tau}^{\prime y},\\&\nonumber
S_{i, \tau}^z=- S_{i, \tau}^{\prime x}\cos\chi + S_{i, \tau}^{\prime z}\sin\chi,
\end{align}
where $\pm$ applies to $\tau=B$ and $\tau=T$ respectively. This rotation does not affect the ferromagnetic term on each layer ($J\to-|J|$). 

In this system both the out-of-plane DMI (${\bf D}=D\bold z$) and the in-plane DMI (${\bf D}=D\bold x$) contribute to linear order in spin wave theory  valid at low temperatures. For $\bold D \parallel \bold B$, the rotation in Eq.~\ref{trans1} rescales the DMI as $D\to D\sin\chi$. On the other hand,  for $\bold D \perp \bold B$,  we have  $D\to D\cos\chi$. However,  the DMI does not contribute to the classical energy given by
\begin{align}
E_{cl}/\mathcal{N}S=-\frac{3}{2}JS- h\sin\chi - \frac{1}{2}J_0S\cos2\chi,
\end{align}
where $\mathcal{N}$ is the total number of sites. Minimizing the classical energy yields that canting angle $\sin\chi= h/h_s$, where $h_s=2J_0S$. The $J$ term is invariant as mentioned above but the DMIs are given by
\begin{align}
&H_{DMI,z}^\tau= D_{z,\chi}\sum_{\la\la i, j\ra\ra}\nu_{ij}\hat{\bold z}\cdot {\bf S}_{i}\times {\bf S}_{j},\\&
H_{DMI,x}^\tau=  \mp D_{x,\chi}\sum_{\la\la i, j\ra\ra}\nu_{ij}\hat{\bold z}\cdot {\bf S}_{i}\times {\bf S}_{j},
\end{align}
where $D_{z,\chi}(D_{x,\chi})=D\sin\chi(\cos\chi)$ and the primes have been dropped. The latter case changes sign on the top and bottom layers as shown in the text. Using the Holstein-Primakoff transformation
\begin{align}
&S_{i}^{ z}= S-b_{i}^\dagger b_{i},\label{hp}\nonumber\\&
 S_{i}^{ y}=  i\sqrt{ S/2}(b_{i}^\dagger -b_{i}),\\&\nonumber
 S_{i}^{ x}=  \sqrt{S/2}(b_{i}^\dagger +b_{i}),
 \end{align}
 the interlayer coupling is given by
\begin{align}
H_{int}&=J_0S\sum_{i\in T, j\in B}\bigg[(n_i+ n_{j})\cos2\chi \nonumber-( b_{i}^\dagger b_{j}+ h.c.)\sin^2\chi\\&\nonumber+( b_{i}^\dagger b_{j}^\dagger+ h.c.)\cos^2\chi\bigg].
\end{align}
It is straight forward to compute the magnon bands. Interestingly, at the saturation field $h=h_s$ one recovers ferromagnetic coupled layers.

For fully antiferromagnetic coupled layers with $J, J_0 >0$, the appropriate rotation
for the bottom layer is given by
\begin{align}
&S_{i,A_1(B_1)}^x=\pm S_{i,A_1(B_1)}^{\prime x}\sin\chi \pm S_{i,A_1(B_1)}^{\prime z}\cos\chi,\label{trans}\nonumber\\&
S_{i,A_1(B_1)}^y=\pm S_{i,A_1(B_1)}^{\prime y},\\&\nonumber
S_{i,A_1(B_1)}^z=- S_{i,A_1(B_1)}^{\prime x}\cos\chi + S_{i,A_1(B_1)}^{\prime z}\sin\chi.
\end{align}
 For the top layer,  we perform the rotation
\begin{align}
&S_{i,A_2(B_2)}^x=\mp S_{i,A_2(B_2)}^{\prime x}\sin\chi \mp S_{i,A_2(B_2)}^{\prime z}\cos\chi,\label{trans}\nonumber\\&
S_{i,A_2(B_2)}^y=\mp S_{i,A_2(B_2)}^{\prime y},\\&\nonumber
S_{i,A_2(B_2)}^z=-S_{i,A_2(B_2)}^{\prime x}\cos\chi + S_{i,A_2(B_2)}^{\prime z}\sin\chi.
\end{align}
The terms that contribute to linear spin wave theory are as follows,
\begin{align}
&H_{J}^{T(B)}= J\sum_{\la i,j\ra}[\cos 2\chi (S_{i,x}S_{j,x}-S_{i,z}S_{j,z})-S_{i,y}S_{j,y}],\\&
H_{DMI,z}^{T(B)}= D_{z,\chi}\sum_{\la\la i, j\ra\ra}\nu_{ij}\hat{\bold z}\cdot {\bf S}_{i}\times {\bf S}_{j},\\&
H_{DMI,x}^{T(B)}= \mp D_{x,\chi}\sum_{\la\la i, j\ra\ra}\hat{\bold z}\cdot {\bf S}_{i}\times {\bf S}_{j},\\&
H_{Z}^{T(B)}= -h\sin{\chi}\sum_i S_{i,z},\\&
H_{J_0}= J_0\sum_{i\in T; j\in B}[\cos 2\chi (S_{i,x}S_{j,x}-S_{i,z}S_{j,z})-S_{i,y}S_{j,y}].
\end{align}
The classical energy is given by
\begin{align}
E_{cl}/\mathcal{N}S=-\frac{3}{2}JS\cos2\chi- h\sin\chi - \frac{1}{2}J_0S\cos2\chi.
\end{align}
 Minimization yields  $h_s=2(3J +J_0)S$.
\section{Bilayer Bose-Hubbard model}
In this appendix, we discuss bilayer honeycomb Bose-Hubbard model, recently studied numerically and shown to exhibit non-zero Berry curvature and bosonic edge states \cite{guo}. The Hamiltonian is given by
\begin{align}
H&= -t\sum_{\langle ij\rangle,\tau}( b^\dg_{i,\tau} b_{j,\tau} +  b^\dg_{j,\tau} b_{i,\tau})  -\sum_{i,\tau} (\mu+U_i) n_{i,\tau}\label{hardcore},\\
H_{int.}&= -t_0\sum_{ i\in T, j\in B}( b^\dg_i b_j +  b^\dg_j b_i), \end{align}
where,  $t>0$ denotes NN hopping, $\mu$ is the chemical potential, and $U_i$ is a staggered on-site potential with $U_i=\Delta$ for $i\in T$ and $U_i=-\Delta$ for $i\in B$,   $n_i=b^\dg_ib_i$ is the occupation number which is either $0$ or $1$.   $b^\dg_i$ and $ b_i$ are the bosonic creation and annihilation operators respectively. They obey the hard-core constraints $[b_i, b_j^\dg]=0$ for $i\neq j$ and $\lbrace b_i, b_i^\dg \rbrace=1$. In the first approximation, only the vertical interlayer coupling  $t_0$ contributes, while $2\Delta$ is the interlayer potential difference. 

 In this system, the DMI is zero (causes a sign problem in QMC) and nonzero Berry curvature is induced by the staggered onsite potential without a  QMC sign problem \cite{guo}.   In order to study the bilayer Bose-Hubbard model  analytically, it is expedient to map  it  to quantum spin Hamiltonian since the bilayer Bose-Hubbard model is a hard-core boson model. This is achieved via the  transformation \cite{matq},  $S_i^+ \to b^\dagger_i,~S_i^-\to b_i,~S_i^z\to n_i-1/2$. The resulting quantum spin Hamiltonian is given by
 \begin{align}
H&=-J\sum_{\langle ij\rangle,\tau}(S_{i,\tau}^+S_{j,\tau}^-+ S_{i,\tau}^-S_{j,\tau}^+)-\sum_{i,\tau}(\mu+U_i)S_{i,\tau}^z,\\
H_{int.}&=-J_0\sum_{i\in T, j\in B}(S_i^+S_j^-+ S_i^-S_j^+),
\label{hh1}
\end{align}
where $S^{\pm}=S^x\pm iS^y$ and $t(t_0)\to J(J_0)$.

 For $J_0=0$, this model reduces to single-layer Bose-Hubbard model on the honeycomb lattice,  where $U_i$  can be considered as a staggered sublattice potential (magnetic field). There are several insulating phases uncovered by QMC phase diagram  \cite{guo}. We have recently complemented the QMC results using the present HP bosonic method \cite{sol2}.  We uncovered the same quantum phase diagram and Berry curvatures seen in QMC  \cite{guo}. Thus, the topological properties of  the single-layer Bose-Hubbard model uncovered by QMC correspond  exactly to  topological properties of magnon bulk bands of the corresponding spin-$1/2$ quantum magnets.  This is due to the mapping  between hard-core bosons and quantum spin systems.  We also noticed that the Berry curvature of each band shows equal and opposite peaks at the corners of the Brillouin zone \cite{sol2}. This means that the integrated Berry curvature vanishes identically for each band, hence thermal Hall effect is not expected to manifest in this system without the DMI.  For the bilayer model  $J_0\neq 0$, the analysis follows a similar procedure. In the HP boson representation, one performs spin wave theory about each insulating phase, then it is straightforward to study the magnetic excitations in each insulating phase similar to single-layer system \cite{sol2}.


\begin{thebibliography}{99}
\bibitem{dm}
 I. Dzyaloshinsky, J. Phys. Chem. Solids {\bf 4}, 241 (1958);  T. Moriya, Phys. Rev. {\bf 120}, 91 (1960).
\bibitem{alex0}
 H. Katsura, N. Nagaosa, and P. A. Lee,   Phys. Rev. Lett.  {\bf 104},  066403 (2010).
  \bibitem{alex1}
Y. Onose, T. Ideue, H. Katsura, Y. Shiomi, N. Nagaosa, Y. Tokura,  Science  { \bf 329}, 297 (2010).
  \bibitem{alex1a}
T. Ideue, Y. Onose, H. Katsura, Y. Shiomi, S. Ishiwata, N. Nagaosa, and Y. Tokura, Phys. Rev. B. {\bf 85}, 134411 (2012).
  \bibitem{zhh} 
  L. Zhang, J. Ren, J. S. Wang, and B. Li, Phys. Rev. B {\bf 87}, 144101 (2013).
  \bibitem{shin}
 R. Shindou et.al., Phys. Rev. B 87,  174427 (2013); Phys. Rev. B 87, 174402 (2013).
  \bibitem{shin1}
R. Matsumoto, R. Shindou, and S. Murakami, Phys. Rev. B 89, 054420 (2014).
\bibitem{yu6}
M. Z. Hasan and C. L. Kane, Rev. Mod. Phys. {\bf 82}, 3045 (2010).
\bibitem{yu7}
X. -L.  Qi and S. -C. Zhang, Rev. Mod. Phys. {\bf 83}, 1057 (2011).

   \bibitem{fdm}
F. D. M. Haldane, Phys. Rev. Lett. {\bf 61}, 2015 (1988).
 \bibitem{alex2}
 R. Matsumoto and S. Murakami, Phys. Rev. Lett. {\bf 106}, 197202 (2011); Phys. Rev. B {\bf 84}, 184406 (2011).
 \bibitem{alex7}
 A.  A. Kovalev and V.  Zyuzin, Phys. Rev. B {\bf 93}, 161106(R) (2016).
 



 \bibitem{thou}
D. J. Thouless, M. Kohmoto, M. P. Nightingale, and M. den Nijs, Phys. Rev. Lett. {\bf 49}, 405  (1982); M. Kohmoto, Annals of Physics {\bf 160}, 343 (1985).
 \bibitem{alex44}
 H. Lee, J. H. Han, and P. A. Lee,   Phys. Rev. B.  {\bf 91},  125413 (2015).
\bibitem{alex4}
A.  Mook, J.  Henk, and I. Mertig, Phys. Rev. B {\bf 90}, 024412 (2014); A.  Mook, J.  Henk, and I. Mertig,  Phys. Rev. B {\bf 89}, 134409 (2014).

 \bibitem{alex6a}
R. Chisnell,  J. S.  Helton, D. E. Freedman, D. K. Singh, R. I.  Bewley, D. G.  Nocera, and Y. S. Lee, Phys. Rev. Lett. {\bf 115}, 147201 (2015).
 \bibitem{alex6}
Max Hirschberger, Robin Chisnell, Young S. Lee, and  N. P. Ong,  Phys. Rev. Lett.  {\bf 115}, 106603 (2015).
  \bibitem{cas} A. H. Castro Neto, F. Guinea, N. M. R. Peres, K. S. Novoselov, and A. K. Geim, Rev. Mod. Phys. {\bf 81}, 109 (2009).
 \bibitem{xc} X. Cao, K. Chen and D.  He, J. Phys.: Condens. Matter {\bf 27}, 166003   (2015).
\bibitem{sol}
S. A.  Owerre, J. Phys.: Condens. Matter {\bf 28}, 386001  (2016). 
\bibitem{sol1}
S. A.  Owerre, J. Appl. Phys. {\bf 120}, 043903 (2016).
\bibitem{kkim}
S. K. Kim, H. Ochoa, R.  Zarzuela, Y.  Tserkovnyak,  arXiv:1603.04827.
\bibitem{mook} A. Mook, J.  Henk, and I. Mertig, Phys. Rev. B {\bf 91}, 224411 (2015).
\bibitem{aat1}
Y. Miura, R.  Hirai, Y.  Kobayashi, and M.  Sato,  J. Phys.
Soc. Jpn. {\bf 75}, 084707 (2006).
\bibitem{aat}
 A. A. Tsirlin, O. Janson, and H. Rosner, Phys. Rev. B 
 {\bf 82}, 144416 (2010);
  \bibitem{aat0}
Y. Singh and P. Gegenwart,  Phys. Rev. B {\bf 82}, 064412 (2010); Phys. Rev. Lett. {\bf 108} 127203 (2012).
  \bibitem{aat00}
X. Liu, T. Berlijn, W.-G. Yin, W. Ku, A. Tsvelik, Young-June Kim, H. Gretarsson, Yogesh Singh, P. Gegenwart, J. P. Hill, Phys. Rev. B {\bf 83}, 220403(R) (2011); S. K. Choi, R. Coldea, A. N. Kolmogorov, T. Lancaster, I. I. Mazin, S. J. Blundell, P. G. Radaelli, Yogesh Singh, P. Gegenwart, K. R. Choi, S.-W. Cheong, P. J. Baker, C. Stock, J. Taylor., Phys. Rev. Lett. {\bf 108}, 127204 (2012).

 \bibitem{HP}
T. Holstein and H. Primakoff, Phys. Rev. {\bf 58}, 1098 (1940).
 \bibitem{mcc}
Ed.  McCann and M.  Koshino, Rep. Prog. Phys. {\bf 76}, 056503 (2013).
\bibitem{mcc1}
E. McCann  and V. I.  Fal'ko , Phys. Rev. Lett.  {\bf 96}, 086805 (2006).
\bibitem{mcc2}
F. Guinea, A. H.  Castro Neto, and N. M. R. Peres, Phys. Rev. B {\bf 73}, 245426  (2006).

\bibitem{mcc3}
J. Nilsson,  A. H.  Castro Neto, F. Guinea, and N. M. R. Peres, Phys. Rev. B {\bf 78}, 045405  (2008).
\bibitem{mak55}
E. Prada, P. San-Jose, L. Brey, and H. A. Fertig, Solid State Commun. 151, 1075 (2011).


 
\bibitem{castt} Eduardo V. Castro, N. M. R. Peres, J. M. B. Lopes dos Santos, A. H. Castro Neto, F. Guinea, Phys. Rev. Lett. {\bf 100}, 026802 (2008).

\bibitem{matq}
T. Matsubara and H. Matsuda,  Prog. Theor. Phys. {\bf 16}, 569 (1956).
\bibitem{guo}
H. Guo, Y. Niu, S. Chen, S. Feng, Phys. Rev. B. {\bf 93}, 121401(R) (2016).
 \bibitem{sol2}
S. A.  Owerre, 	arXiv:1603.07989 (accepted, J. Phys.: Condens. Matter).

\bibitem{jott}
G. Jotzu, M. Messer, R. Desbuquois, M. Lebrat, Th. Uehlinger, D. Greif, T. Esslinger, Nature {\bf 515}, 237 (2014).
\bibitem{zhe}
Z. Qiao, W. -K. Tse, H. Jiang, Y. Yao, Q. Niu, Phys. Rev. Lett.  {\bf 107}, 256801 (2011).
\bibitem{ll}
Patrik Fazekas, Lectures Notes on Electrons Correlations and Magnetization (World Scientific Publishing Co. Pte. Ltd, 1999); p. 283-285.

\end{thebibliography}
\end{document}